\documentclass[twocolumn,aps, amsfonts, amsmath,superscriptaddress, nofootinbib, showpacs,byrevtex, floatfix,preprintnumbers,prd]{revtex4}    
\usepackage{subfigure} 
\usepackage{amsmath,amssymb} 
\usepackage{graphicx} 
\usepackage{rotating} 
\usepackage{bm} 
\usepackage{color} 
\usepackage{hyperref} 

\newcommand{\f}{\mathbf} 
\def\fp{\f{p}} 
\def\fq{\f{q}}

\def\fx{\f{x}}

\def\fl{\boldsymbol{\ell}} 
\def\s{\sigma} 
\def\sb{\bar{\sigma}} 
\def\cb{\bar{c}} 
\def\D{\mathcal{D}} 
\def\lf{\left} 
\def\ri{\right} 
 
\def\dl{\delta} 
\def\fc{\frac} 
\newcommand{\lfc}[2]{{\textstyle\frac{#1}{#2}}} 
\def\Tr{\mbox{Tr}} 
\def\STr{\mbox{STr}} 
\def\G{\Gamma} 
\def\om{\omega} 
\def\df{\dfrac}

\def\be{\begin{equation}} 
\def\ee{\end{equation}} 
\def\b*{\begin{equation*}} 
\def\e*{\end{equation*}} 

\newcommand{\hk}{\hspace{0.1cm}} 
\newcommand{\lk}{\left(} 
\newcommand{\rk}{\right)} 
\newcommand{\bea}{\begin{eqnarray}} 
 \newcommand{\eea}{\end{eqnarray}}

\def\di{\displaystyle} 
 
\def\bg{\begin{eqnarray}\begin{array}{rcl}\displaystyle} 
\def\eg{\end{array} &\di    &\di   \end{eqnarray}} 
\def\bm#1{\begin{eqnarray}\begin{array}{#1}\di} 
\def\bmo#1{\begin{eqnarray*}\begin{array}{#1}\di} 
\def\bml#1#2{\begin{eqnarray}\begin{array}{#1}\label{#2}\di} 
\def\bgo{\begin{eqnarray*}\begin{array}{rcl}\displaystyle} 
\def\ego{\end{array} &\di    &\di \nonumber  \end{eqnarray*}} 
\def\btensor#1#2{\renew\left#1\begin{array}{#2}\di} 
\def\brtensor#1#2#3{\ren#3\left#1\begin{array}{#2}} 
\def\botensor#1#2{\renew\left#1\begin{array}{#2}} 
\def\etensor#1{\end{array}\right#1} 
 
\def\eq#1{(\ref{#1})} 
\def\Eq#1{Eq.~(\ref{#1})} 
 
\def\0#1#2{\frac{#1}{#2}} 
 
\def\e{\mbox{\boldmath$\epsilon$}} 

\def\ren#1{\renewcommand{\arraystretch}{#1}} 
 
\def\renew{\renewcommand{\arraystretch}{1}} 

\makeatletter					
\renewcommand*\env@matrix[1][\arraystretch]{%
  \edef\arraystretch{#1}%
  \hskip -\arraycolsep 
  \let\@ifnextchar\new@ifnextchar 
  \array{*\c@MaxMatrixCols c}} 
\makeatother

\begin{document} 
 
\title{Hamiltonian Flow in Coulomb Gauge Yang-Mills Theory} 
\date{\today} 
\author{Markus Leder} 
\affiliation{Institut f\"ur Theoretische Physik, Universit\"at T\"ubingen, Auf der Morgenstelle 14, 72076 T\"ubingen, Germany} 
\author{Jan M. Pawlowski} 
\affiliation{Institut f\"ur Theoretische Physik, Universit\"at Heidelberg, Philosophenweg 16, 62910 Heidelberg, Germany} 
\affiliation{ExtreMe Matter Institute EMMI, GSI Helmholtzzentrum f\"ur Schwerionenforschung, \\ Planckstr. 1, 64291 Darmstadt, Germany} 
\author{Hugo Reinhardt} 
\affiliation{Institut f\"ur Theoretische Physik, Universit\"at T\"ubingen, Auf der Morgenstelle 14, 72076 T\"ubingen, Germany} 
\author{Axel Weber} 
\affiliation{Instituto de F\'isica y Matem\'aticas, Universidad Michoacana de San Nicol\'as de Hidalgo, Edificio C-3, Ciudad Universitaria, 58040 Morelia, Michoac\'an, Mexico} 

\pacs{12.38.Aw, 05.10.Cc, 11.10.Ef, 11.15.Tk}

\begin{abstract} 
  We derive a new functional renormalization group equation for
  Hamiltonian Yang-Mills theory in Coulomb gauge. The flow equations
  for the static gluon and ghost propagators are solved under the
  assumption of ghost dominance within different diagrammatic
  approximations. The results are compared to those obtained in the variational approach and the reliability of the approximations
  is discussed.
\end{abstract} 
 
\maketitle 
 
\section{Introduction} 
QCD is experimentally well-tested as the theory of strong interactions
in the perturbative regime where asymptotic freedom holds. In the
non-perturbative regime we have collected much theoretical evidence
ranging from a linearly rising potential for heavy quarks to hadronic
observables which match at least qualitatively the experimental
values.  These non-perturbative results were either obtained by
lattice methods, QCD-model computations and in recent years, to an
increasing extent, by first principle continuum QCD computations based
on functional methods. Despite the impressive success of these
combined efforts there still remains much to be understood,
both qualitatively and quantitatively. Open physics questions range
from the mechanism of confinement and its relation to spontaneous chiral
symmetry breaking to the properties of QCD at finite
temperature and density. Progress in these directions can only be
obtained with a combination of the methods at hand, in particular,
when it comes to the understanding of the underlying physics mechanisms. 

This has triggered an increasing interest in non-perturbative studies
of continuum Yang-Mills theory and full QCD in recent years. Many of
these studies were carried out in Landau gauge, using Dyson-Schwinger
equations (DSE),
for reviews see
\cite{Alkofer:2000wg,Fischer:2006ub,vonSmekal:2008ws,Binosi:2009qm,Fischer:2008uz} (see also \cite{Boucaud:2008ky}),
as well as the functional renormalization group (FRG) equations,
for reviews see
\cite{Litim:1998nf,Pawlowski:2005xe,Gies:2006wv} (see also \cite{Ellwanger:1995qf}). Another attractive
possibility is Coulomb gauge, which has been pursued either with DSEs,
e.g. {\cite{Zwanziger:1998ez,Watson:2006yq,Alkofer:2009dm}}, or with a
variational approach to the Hamiltonian formulation
{\cite{Szczepaniak:2001rg,Feuchter:2004mk,Reinhardt:2004mm,Schleifenbaum:2006bq,Epple:2006hv,Epple:2007ut}},
for a short introduction see \cite{Reinhardt:2008ax}. Each approach
has its own advantages and drawbacks and by combining the different
approaches one can expect to gain new insights into the theory, in
particular, into the non-perturbative regime.

In this spirit we put forward in the present paper a functional
renormalization group approach to the Hamiltonian formulation of
Yang-Mills theory. A specific advantage of the Hamiltonian formulation
is its close connection to physics as demonstrated in various
applications, see e.g. \cite{Reinhardt:2008ek}. However, Hamiltonian renormalization is a well-known
non-trivial task which has hampered progress in the Hamiltonian
approach for many years. Specifically it complicates the search for
RG-invariant, self-consistent approximations to the full vacuum wave
functional. In the present Hamiltonian FRG approach to Coulomb gauge
Yang-Mills theory the latter task is directly solved by a
self-consistent approximation to the flow equation. Such a procedure
guarantees RG invariance by its very definition, and hence combines
the advantages of the Hamiltonian approach with that of the FRG.  We
will also show explicitly that with suitable truncations the
integrated flow equations become precisely the DSEs of the
variational approach \cite{Feuchter:2004mk} with a Gaussian ansatz for
the vacuum wave functional. In our explicit computations we will focus
on the infrared sector of the theory within the ghost-dominance
scenario. Therefore we only include full momentum-dependent gluon and
ghost propagators and the bare ghost-gluon vertex. The latter
approximation is justified by the non-renormalization of the full
ghost-gluon vertex.

The paper is organized as follows: In section II we present the
basics of the functional renormalization group (FRG) flow equation
approach. In particular, we derive the flow equation for the
Hamiltonian approach to Yang-Mills theory in Coulomb gauge. The gauge
fixing in the scalar product of the wave functional is accomplished
by the Faddeev-Popov method. In section III the FRG flow equations for
the gluon and ghost propagators are derived assuming a bare,
non-running, ghost-gluon vertex. We also extend the uniqueness proof
for the infrared scaling solution in Landau gauge to Coulomb gauge
Yang-Mills theory. In sect. IV we numerically solve the flow equations for
ghost and gluon propagators within two different approximations which
are compared with the DSE results. Finally our conclusions are given
in sect. V. Some mathematical details of the derivation of the flow
equations of the Hamiltonian approach as well as some details of the
numerical procedure are deferred to the appendices.
 
\section{Derivation of the Hamiltonian flows} 
 
\subsection{Flow equation for the effective action} 
 
Below we briefly summarize the essential ingredients of the FRG 
approach. The starting point is the finite renormalized generating 
functional of the Green's functions, 
\begin{equation} 
\label{4-1} 
Z [j] = \int [D \varphi]^{\ }_{\rm ren} e^{- S [\varphi] + j \cdot \varphi}  \,, 
\end{equation} 
where the subscript ``ren'' of the measure indicates an 
appropriate renormalization procedure that renders the functional 
integral in (\ref{4-1}) finite, for more details see 
\cite{Pawlowski:2005xe}. Here $\varphi$ and $j$ denote collectively all 
fields involved and the corresponding sources. Furthermore, the 
scalar product $j \cdot \varphi$ includes summation over all 
indices and integration over space-time. Finally, the theory is 
specified by the classical action $S [\varphi]$ and the renormalized functional 
integral measure. In the functional renormalization group (FRG) approach the 
generating functional is IR regularized by adding a regulator term 
$\Delta S_k$ to the classical action $S$. It depends on a cut-off 
momentum $k$ and is chosen to be quadratic in the fields, 
\begin{equation} 
\label{4-2} 
\Delta S_k [\varphi]  =  \frac{1}{2} \varphi \cdot R_k \cdot \varphi  
 \equiv  \frac{1}{2} \int \frac{d^d p}{(2 \pi)^d} \varphi (-p) R_k (p) \varphi (p) \hk . 
\end{equation} 
The regulator function $R_k (p)$ is an effective momentum dependent  
mass and has the properties  
\begin{eqnarray}\label{regulator properties} 
\lim\limits_{p/k \to 0} R_k (p) & > & 0 \nonumber\\ 
\lim\limits_{k/p \to 0} R_k (p) & = & 0 \hk . 
\end{eqnarray} 
The first condition ensures that $R_k (p)$ is indeed an infrared 
regulator and suppresses the propagation of the modes with momentum $p 
\lesssim k$. The second condition implies that the momentum modes with 
$p \gg k$ are unaffected by the regulator and that the full finite 
renormalized generating functional of the theory at hand is recovered 
as the cut-off scale $k$ is pushed to zero. 
 
\begin{figure*}[t]
{\large$k\partial_k\quad$} \parbox{50pt}{\includegraphics[width=50pt,clip=]{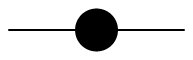}}{\large$\;^{-1}\quad=\quad$} 
\parbox{90pt}{\includegraphics[width=90pt,clip=]{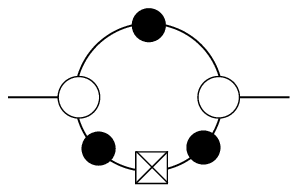}}
{\large$\quad+\quad$} \parbox{90pt}{\includegraphics[width=90pt,clip=]{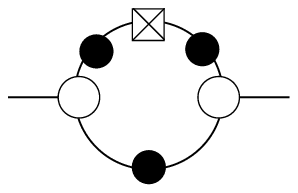}} 
{\large$\quad-\quad$}\parbox{65pt}{\includegraphics[width=65pt,clip=]{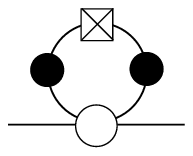}}
\caption{Flow equation for the inverse propagator (2-point function) 
  $\Gamma_k^{(2)} = \langle \phi \phi \rangle^{- 1}$. Here and in all the diagrammatic equations below, lines with full 
  circles represent  
  $\lk \Gamma_k^{(2)} + R_k \rk^{- 1}$ on the r.h.s. of the equation, but only ${\Gamma_k^{(2)}}^{- 1}$ on its l.h.s. Open circles represent full 
  proper vertices $\Gamma_k^{(n)}$, while the square with a cross 
  represents the regulator insertion $\dot{R}_k$.} 
\label{fig4-1} 
\end{figure*}

The basic idea of the FRG flow equation approach to Yang-Mills theory 
is to start at a large cut-off scale $k$, where the theory is under 
control due to asymptotic freedom, and then let the cut-off $k$ flow 
to the small momentum regime, which is non-perturbative. The evolution 
of the Green's functions with the cut-off scale $k$ is described by a 
flow equation which is obtained by taking the derivative of the regulated 
generating functional 
\begin{equation} 
\label{4-4} 
Z_k [j] = \int [D \varphi]^{\ }_{\rm ren}   
e^{- S [\varphi] - \Delta S_k [\varphi] + j \cdot \varphi} \equiv e^{W_k [j]} 
\end{equation} 
w.r.t. the momentum scale $k$. We also remark that the 
finiteness of $Z_k[j]$ follows from that of $Z[j]$ with the 
representation  
\begin{equation}\label{eq:finite} 
Z_k[j]=e^{-\Delta S_k[\delta/\delta j]}\, Z[j]\, 
\end{equation}  
and the differentiability of $Z[j]$ w.r.t. $j$. With \eq{eq:finite} the flow of $Z_k$ is 
derived as 
\begin{equation} 
\label{4*} 
\partial_t Z_k [j] = \lk - \frac{1}{2} \frac{\delta}{\delta j}  
\cdot \dot{R}_k \cdot \frac{\delta}{\delta j} \rk Z_k [j] \hk , 
\end{equation} 
where the dot on $R$ stands for the derivative w.r.t. the 
dimensionless variable $t = \ln k /k_0$. Here $k_0$ is an arbitrary 
reference scale. The functional derivative $\delta/\delta j$ in  
momentum space is to be understood as 
\begin{equation} 
\frac{\delta}{\delta j} \, (p) \equiv (2 \pi)^d \, \frac{\delta}{\delta j (-p)}  
\hk . 
\end{equation} 
Expressing (\ref{4*}) in terms of the generating functional of the  
connected Green's functions, $W_k [j]$, defined in 
(\ref{4-4}), we get  
\begin{equation} 
\label{4-X1} 
\partial_t W_k [j] = - \frac{1}{2} \frac{\delta W_k}{\delta j}  
\cdot \dot{R}_k \cdot \frac{\delta W_k}{\delta j} - \frac{1}{2}  
\Tr \, \dot{R}_k \frac{\delta^2 W_k}{\delta j \delta j}  \hk , 
\end{equation} 
where 
\begin{equation} 
\Tr \, \dot{R}_k \frac{\delta^2 W_k}{\delta j \delta j} \equiv 
\int \frac{d^d p}{(2 \pi)^d} \, \dot{R}_k (p) \, (2 \pi)^{2d} 
\frac{\delta^2 W_k}{\delta j (-p) \delta j (p)} \hk . 
\end{equation} 
By taking derivatives of \eq{4-X1} w.r.t. $j$ one obtains the flow 
equations for the connected Green's functions.  In practice, it is 
usually more convenient to perform first a Legendre transform from the 
sources $j$ to the classical field  
\begin{equation} 
\label{4-5} 
\phi = \frac{\delta W_k [j]}{\delta j} \, , 
\end{equation} 
resulting in the effective action 
\begin{equation} 
\label{4-6} 
\Gamma_k [\phi] = \lk - W_k [j] + j \cdot \phi \rk_{j = j_k 
  (\phi)} - \frac{1}{2} \phi\cdot  R_k \cdot \phi \, , 
\end{equation} 
where $j_k (\phi)$ is given by solving (\ref{4-5}) for $j$. Hence, \eq{4-6} also implies that  
\begin{equation} 
\label{Gamma1} 
j = \frac{\delta( \Gamma_k +\Delta S_k)}{\delta \phi} \, .  
\end{equation} 
By taking the derivative of the effective action $\Gamma_k$ w.r.t. $t = \ln k / k_0$ and using its definition \eq{4-6} and 
(\ref{4-X1}) one arrives at 
\begin{equation} 
\label{4-X2} 
\partial_t \Gamma_k [\phi] = \frac{1}{2} \Tr \, \dot{R}_k  
\frac{\delta^2 W_k}{\delta j\delta j} \hk . 
\end{equation} 
The second derivative of $W_k$ w.r.t. the currents is the connected 
two-point function, the propagator of the theory. It is related to the 
inverse of the second derivative of the effective action,  
\begin{equation} 
\label{5-11} 
\frac{\delta^2 W_k}{\delta j \delta j} = \lk \frac{\delta^2 
  \Gamma_k}{\delta \phi \delta \phi} + R_k \rk^{- 1} \, .   
\end{equation} 
\Eq{5-11} follows directly from \eq{Gamma1} and 
\begin{equation} 
  \frac{\delta}{\delta \phi} \, j =  
  \lk \frac{\delta}{\delta  j} \, \phi \rk^{- 1} =  
  \lk \frac{\delta^2 W_k}{\delta j \delta j} \rk^{- 1} \,.  
\end{equation} 
Inserting \eq{5-11} into (\ref{4-X2}) we obtain Wetterich's flow 
equation for the effective action \cite{Wetterich:1992yh}, 
\begin{equation} 
\label{4-7} 
\dot{\Gamma}_k [\phi] = \frac{1}{2} \Tr \, \dot{R}_k  
\lk \Gamma^{(2)}_k [\phi] + R_k \rk^{-1} \,, 
\end{equation} 
where the dot denotes the derivative w.r.t. $t$ and  
\begin{equation} 
\label{4-8} 
\Gamma^{(n)}_{k, 1\cdots n} [\phi] = \frac{\delta^n \Gamma_k 
  [\phi]}{\delta \phi_1 \cdots \delta \phi_n}  
\end{equation}  
are the one-particle irreducible $n$-point functions (proper 
vertices). We have also introduced a condensed notation where $n$ stands 
for the space-time variable, $\phi_n=\phi(x_n)$. The generic structure 
of this equation is independent of the details of the underlying 
theory, i.e., of the explicit form of the action $S [\phi]$, but is a 
mere consequence of the form of the regulator term (\ref{4-2}), i.e., that it is quadratic in the fields.  By taking functional derivatives of 
Eq. (\ref{4-7}) w.r.t. the fields one obtains the flow 
equations for the (inverse) propagators or vertices. Taking the second 
functional derivative we find the flow equation for the two-point 
function, 
\begin{widetext} 
\begin{equation} 
\label{F5A-X3} 
\dot{\Gamma}^{(2)}_{k,12} = \frac{1}{2} \Tr \, 
  \dot{R}_k \lk \Gamma^{(2)}_k [\phi] + R_k \rk^{-1} 
\left(-\Gamma^{(4)}_{k,12} 
  + \left[\Gamma^{(3)}_{k,1}  
    \lk \Gamma^{(2)}_k [\phi] + R_k \rk^{-1} \Gamma^{(3)}_{k,2}  
    +(1 \leftrightarrow 2) \right]\right) 
  \lk \Gamma^{(2)}_k [\phi] + R_k \rk^{-1} \,. 
\end{equation} 
\end{widetext} 
In (\ref{F5A-X3}) we have suppressed all cyclic indices, which are 
summed (integrated) over in the trace. The flow equation 
(\ref{F5A-X3}) for the propagator is diagrammatically illustrated in 
Fig. \ref{fig4-1}.

\subsection{Hamiltonian flows}\label{sec:hamflow} 

In this section we derive the flow equation for the effective action in the Hamiltonian approach. Some details can be found in Appendix \ref{app details deriv act}.
In the Hamiltonian approach the generating functional of the static 
(equal-time) Green's functions is defined by \begin{equation} 
\label{5-9} 
Z [j] = \langle \psi | \exp (j \cdot \varphi) | \psi \rangle = \int \D 
\varphi | \psi [\varphi] |^2 \exp (j \cdot \varphi) \hk ,  
\end{equation}  
where $\langle \varphi | \psi \rangle = \psi [\varphi] $ is the vacuum wave 
functional.  
Comparison with Eq. (\ref{4-1}) shows that in the 
Hamiltonian approach the ``action'' is defined by the wave functional 
\begin{equation}\label{eq:wavephi} 
\exp \lk - S [\varphi] \rk \equiv | \psi [\varphi] |^2 \,. 
\end{equation} 
With this identification Eq. (\ref{5-9}) has precisely the standard 
form of the generating functional (\ref{4-1}), except that the 
functional integral extends over time-independent fields, and we can 
repeat, step by step, the formal manipulations of the last section to 
derive the corresponding FRG flow equation, which has the same 
structure as Eq. (\ref{4-7}). 
 
In the present work we are specifically interested in the Hamiltonian flow of Yang-Mills theory in Coulomb gauge. In this case 
$\varphi$ stands for the transverse spatial components of the gauge field $A$, 
\begin{equation}\label{eq:transverse}  
t_{ij} A_j=A_i \quad {\rm with} \quad  t_{ij}(\fp) = 
    \delta_{ij} - \frac{p_i p_j}{\fp^2}\,. 
\end{equation} 
Note that in the present Hamiltonian context, we denote the  
contravariant components of all 3-vectors with subscripts. 
The wave functional $\psi(A)$ is the true Yang-Mills vacuum wave 
functional restricted to transverse fields.  Implementing the Coulomb 
gauge in the standard fashion by the Faddeev-Popov method, the 
generating functional (\ref{5-9}) reads 
\begin{equation}\label{6-11}
Z [J] = \int \D A \det (- \partial D) | \psi [A]|^2 \exp (J \cdot A) \hk , 
\end{equation} 
where $\det (- \partial D)$ denotes the Faddeev-Popov determinant and 
\begin{equation}\label{eqcovder} 
D^{ab} = \delta^{ab} \partial + g f^{acb} A^c\,, 
\end{equation} 
is the covariant derivative in the adjoint representation. The short 
hand notation $J\cdot A$ also includes the internal  
indices, 
\begin{equation} 
J \cdot A  =  \int \frac{d^3 p}{(2 \pi)^3} J^a_i (-\fp) A^a_i (\fp) \, . 
\end{equation} 
The Faddeev-Popov determinant $\det (- \partial D)$ is 
represented by an integral over ghost fields $c^a (\fx), \bar{c}^a 
(\fx)$. After introducing regulator terms for ghosts and gluons via 
$\Delta S_k$, see Eq. (\ref{4-2}), the regulated generating functional (\ref{4-4}) of the 
Hamiltonian approach to Yang-Mills theory in Coulomb gauge reads  
\begin{eqnarray}\nonumber  
\label{5-13} 
Z_k [J, \sigma, \bar{\sigma}] &=& \int \D A \D \bar{c} \D c \,  
e^{- S - \Delta S_k  + J \cdot A + \bar{\sigma} \cdot c + \bar{c} \cdot \sigma } \\[1ex] 
&\equiv &\exp  W_k [J, \sigma, \bar{\sigma}] \,. 
\end{eqnarray} 
The ``classical action'' $S$ in \eq{5-13} is given by   
\begin{equation} 
\label{6-14} 
S = - \ln | \psi [A] |^2 + \int d^3 x \,\bar{c}^a (\fx) (- \partial D)^{ab} c^b (\fx) 
\end{equation} 
and the regulator term is chosen as  
\begin{equation}\label{regulator term} 
\Delta S_k[A,c,\cb] = \lfc{1}{2} A\cdot R^{\ }_{A,k}\cdot A + \cb\cdot R^{\ }_{c,k}\cdot c \,. 
\end{equation} 
Here we have used again the short hand notation for scalar products, e.g.\  
\begin{equation}\label{eq:RkA} 
    A\cdot R_{A,k}^{\ }\cdot A = \int\frac{d^3 p}{(2\pi)^3} 
    A_i^a(-\fp)R_{A,k}^{ab}(\fp)A^b_i(\fp) \,. 
\end{equation} 
In \eq{eq:RkA} the transversality of the gauge field, 
$A^a_i(\fp)=t_{ij}(\fp) A^a_j(\fp)$, is implied. Put differently, we could have 
multiplied the regulator with the transverse projector. Due to 
global color symmetry and spatial rotational symmetry, 
the regulators take the form 
\begin{equation}\begin{split}\label{regulator} 
    R_{A,k}^{ab}(\fp) =& R^{\ }_{A,k}(p) \delta^{ab} \;,\\ 
    R_{c,k}^{ab}(\fp) =& R^{\ }_{c,k}(p) \delta^{ab} \;\;, 
\end{split}\end{equation} 
with the notation $p = |\fp|$ that we shall use from now on. 
Both regulators are chosen to depend on the same 
dimensionless shape function $r_k (p)$. Accounting for dimensions we 
put 
\begin{equation}\begin{split} 
R_{A,k}(p) =&\, 2p\, r_k(p) \,, \\ 
R_{c,k}(p) =&\, g \,p^2\, r_k(p) = g \bar{R}_{c,k}(p) \,. 
\end{split}\end{equation} 
(see our remarks following Eq. (\ref{eq:GkC}) concerning the factor $g$ included in the
definition of $R_{c,k}(p)$). In the numerical solution we have chosen the $r_k (p)$ as given in \eq{explicit shape}.  
 
From the regularized generating functional (\ref{5-13}) one derives 
the flow equation for the effective action as outlined in the previous 
section for a single field. In the present case, for compactness of 
the notation, it is convenient to combine gluon and ghost fields into 
a superfield 
\begin{equation}\label{superfields} 
\varphi = (A, c, \bar{c}) \hk , \hk \bar{\varphi} = (A, - \bar{c}, c) \hk . 
\end{equation} 
Accordingly, we introduce the supersources 
\begin{equation} 
I = (J, \sigma, \bar{\sigma}) \hk , \hk \bar{I} = (J, - \bar{\sigma}, \sigma) 
\end{equation} 
and supermatrices  
\begin{eqnarray} 
\mathcal{R}_k & = & {\rm diag} \lk R_{A,k}^{\ }, R_{c,k}^{\ }, R^T_{c,k} \rk \nonumber\\ 
M & = & {\rm diag} (\mathbf{1}, \mathbf{-1},\mathbf{-1}) \, ,  
\end{eqnarray}  
where $M$ figures as metric in the superspace and enters the definition of the 
supertrace, 
\begin{equation} 
\STr (\dots) \equiv \Tr (M \dots) \hk . 
\end{equation} 
With this notation the effective action (\ref{4-6}) is given by 
\begin{equation} 
\label{F6-*} 
\Gamma_k [\phi] = - W_k [I_k] + I_k \cdot \bar{\phi} - 
\frac{1}{2} \bar{\phi} \cdot \mathcal{R}_k M \cdot \phi \hk , 
\end{equation} 
where $W_k [I_k]$ is defined by Eq. (\ref{5-13}) and 
\begin{equation} 
\label{F6-**}  
\bar\phi = \frac{\delta W_k [I_k]}{\delta I_k} 
\end{equation} 
is the classical superfield $\phi=\langle \varphi\rangle$ with 
$\phi=(A\,,\,c\,,\,\bar c)$, where in a slight abuse of notation we 
use the same symbols for the components of $\phi$ and $\varphi$. The 
flow of the effective action reads (cf. Eq. (\ref{4-X2})) 
\begin{equation}\label{flow gamma with W}
  \partial_t \Gamma_k [\phi] = \frac{1}{2} \STr \,  
  \dot{\mathcal{R}}_k \frac{\delta^2 W_k}{\delta \bar{I} \delta I} \hk . 
\end{equation} 
By means of (\ref{F6-**}) one derives from (\ref{F6-*}) 
(cf. Eq. (\ref{5-11})) 
\begin{equation}\label{inversion relation} 
  \frac{\delta^2 W_k}{\delta \bar{I} \delta I} = \lk  
  \frac{\delta^2 \Gamma_k}{\delta \bar{\phi} \delta \phi} + \mathcal{R}_k \rk^{- 1} 
\end{equation} 
and obtains the desired flow equation (cf. Eq. (\ref{4-7})) 
\begin{equation} 
\label{F6-***} 
\partial_t\Gamma_k = \frac{1}{2} \STr \, \dot{{\cal R}}_k 
\lk \Gamma_k^{(2)}+ {\cal R}_k \rk^{-1} \ \,,  
\end{equation} 
where $\Gamma_k^{(2)}=\delta^2 \Gamma_k/\delta \bar{\phi} \delta 
\phi$. In the present Hamiltonian approach to Yang-Mills theory in 
Coulomb gauge the effective action $\Gamma_k [\phi]$ defined by 
Eqs. (\ref{5-13}), (\ref{6-14}) and (\ref{F6-*}) is exclusively 
determined by the vacuum wave functional $\psi [A]$ and the 
Faddeev-Popov determinant. Importantly, the FRG 
approach does not require the knowledge of the full vacuum wave 
functional. It is sufficient to know the wave functional in the 
asymptotic region $k \to \infty$, where perturbation theory applies. 
The full quantum effective action $\Gamma_{k \to 0}$ and hence the 
full vacuum wave functional is then computed by solving the flow equation, 
making suitable ans\"atze and truncations for $\Gamma_k$ or its derivatives.

\section{Flows for Coulomb gauge Yang Mills Theory} 
 
\subsection{Uniqueness of IR scaling}\label{sec:scaling}  
 
The Hamiltonian flow equation set up for Coulomb gauge Yang-Mills 
theory in the last section~\ref{sec:hamflow} very much resembles the one 
in Landau gauge, but in one dimension less. It has already been 
speculated that there is a close connection between these two 
formulations, \cite{Burgio:2009xp}. Here we shall employ the similarities in order 
to derive unique scaling laws for the infrared behavior of Coulomb 
gauge Yang-Mills theory. 
 
It has been shown in \cite{Fischer:2009tn,Alkofer:2008jy} that Landau 
gauge Yang-Mills theory admits a unique infrared scaling 
solution \cite{Lerche:2002ep}. Moreover, this solution implies ghost dominance. We 
emphasize that uniqueness refers to the unique scaling relations if 
scaling is present. Indeed, Landau gauge Yang-Mills theory also admits a 
solution without such a scaling behavior, the decoupling 
solution. More details can be found e.g. in \cite{Fischer:2008uz}. 
The scaling and decoupling solutions also exist for the DSE obtained 
in the Hamiltonian formulation of Yang-Mills theory in Coulomb gauge 
and were baptized ``critical'' and ``subcritical'' solutions 
\cite{Epple:2007ut}. Furthermore, lattice calculations 
\cite{Burgio:2009xp} show that the scaling or critical solution is 
realized in Coulomb gauge. 
 
The proof in \cite{Fischer:2009tn} was based on the 
comparison of the full hierarchies of DSE and FRG equations for Green's 
functions. Apart from this it only relied on the details of the 
coupling between ghosts and gluons and the canonical scaling 
properties of the gluonic self-coupling. This has been made 
transparent in \cite{Fischer:2009tn}. The proof as formulated there 
can be directly transferred to Coulomb gauge, the only missing piece 
is provided by the flow equation derived in the present work.  
 
With the Hamiltonian Coulomb gauge DSEs and the FRGs we derive the 
same set of constraint equations for the scaling coefficients as in 
\cite{Fischer:2009tn}. There are additional terms coming from the 
higher classical gluonic $n$-point vertices which all can be proven to 
be sub-leading. This relates to the fact that the canonical scaling of 
classical gluonic vertices is less divergent than that of the dressed 
vertices. In summary we conclude that also Coulomb gauge in its 
Hamiltonian formulation admits a unique scaling solution with the same 
scaling laws that are satisfied in Landau gauge in $d=3$. The scaling 
relations relevant for the present work are that for the propagators,  
\begin{equation}\label{eq:scalprops}  
\langle A(p) A(-p)\rangle \propto \0{1}{p^{2(1+\kappa_A)}}\,,\quad   
\langle c(p) \bar c(-p)\rangle \propto \0{1}{p^{2(1+\kappa_c)}}\,,  
\end{equation}  
and for the ghost gluon vertex at the symmetric point,  
\begin{equation} \label{eq:scalghostgluon} 
\fc{\dl^3\G_k}{\dl \cb^a\dl c^b\dl A^c_i} \propto p^{2 \kappa_{\bar c c A}}  
\fc{\dl^3 S}{\dl \cb^a\dl c^b\dl A^c_i} \,. 
\end{equation} 
The scaling solution entails the non-renormalization of the 
ghost-gluon vertex, $\kappa_{\bar c c A}=0$, and a scaling relation 
for the scaling of the ghost and gluon propagator, summarized as
\begin{equation}\label{eq:landau3} 
  \kappa_{\bar c c A}=0 \quad {\rm and}\quad  \kappa_A= -\012 -2 \kappa_c \,,
  \quad {\rm with}\quad \kappa_A\leq -\frac{1}{4}\,.  
\end{equation}  
\Eq{eq:landau3} implies ghost-dominance in the sense that diagrams
with ghost lines dominate in the infrared over diagrams with gluonic
lines, see also \cite{Fischer:2009tn}. The scaling coefficients
$\alpha,\beta$ used in Coulomb gauge are defined via
\begin{equation}\label{eq:scalpropsCoul}  
\langle A(p) A(-p)\rangle \propto p^{\alpha}\,,\quad   
\langle c(p) \bar c(-p)\rangle \propto \0{1}{p^{2+\beta}}\,,  
\end{equation}  
see Eq. (\ref{def alpha beta}) below. The coefficients $\alpha$ and 
$\beta$ relate to the $\kappa$'s via $\alpha=-2 - 2 
\kappa_A$ and $\beta=2 \kappa_c$.  Hence we find {\it unique} scaling 
laws in Coulomb gauge with the scaling relation  
\begin{equation}\label{eq:Cscaling} 
\alpha=2 \beta -1\,.  
\end{equation}  
The sum rule \eq{eq:Cscaling} has been found in DSE analyses in 
Coulomb gauge before \cite{Schleifenbaum:2006bq}, here we have proven its uniqueness.

\subsection{Propagator flows} 
 
\begin{figure*}[t] 
\Large{$k\partial_k$} \parbox{50pt}{\includegraphics[width=50pt,clip=]{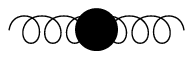}}{\Large$^{-1}=$}
\parbox{90pt}{\includegraphics[width=90pt,clip=]{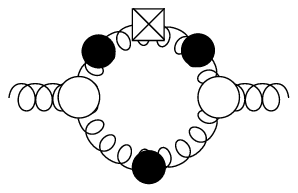}}
{\Large$\;-\;$}\parbox{90pt}{\includegraphics[width=90pt,clip=]{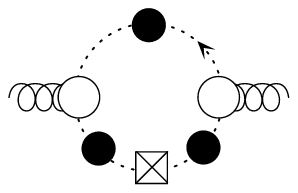}} 
{\Large$\;-\;$}\parbox{90pt}{\includegraphics[width=90pt,clip=]{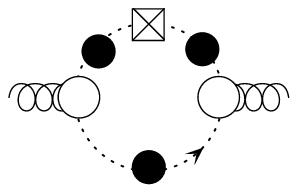}} \\
{\Large$\;-\;$}\parbox{65pt}{\includegraphics[width=65pt,clip=]{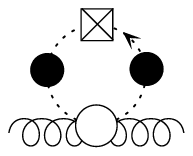}} 
{\Large$\;-\fc{1}{2}\;$}\parbox{65pt}{\includegraphics[width=65pt,clip=]{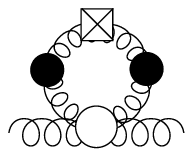}} 
\caption{Flow equation of the gluon propagator, Eq. (\ref{full gluon flow 
    formula}). Here and in the following, the spiral and dotted lines 
  with the black circles denote the regularized gluon and ghost 
  propagators at cutoff momentum $k$, respectively. White circles 
  stand for proper vertices at cutoff $k$, a regulator insertion 
  $\dot{R}_k$ is represented by a square with a cross.} 
\label{gluon full flow}
\end{figure*}

\begin{figure*}[t] 
{\Large$k\partial_k$} \parbox{50pt}{\includegraphics[width=50pt,clip=]{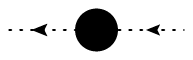}}{\Large$^{-1}=$}
\parbox{90pt}{\includegraphics[width=90pt,clip=]{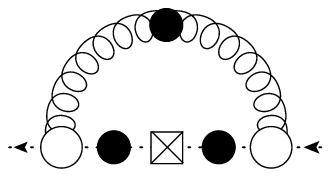}}
{\Large$\;+\;$}\parbox{90pt}{\includegraphics[width=90pt,clip=]{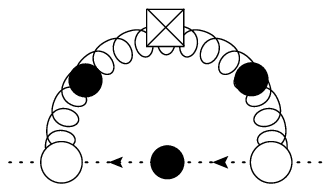}}
{\Large$\;-\fc{1}{2}\;$}\parbox{65pt}{\includegraphics[width=65pt,clip=]{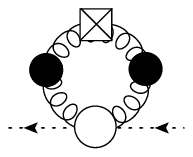}}
{\Large$\;-\;$}\parbox{65pt}{\includegraphics[width=65pt,clip=]{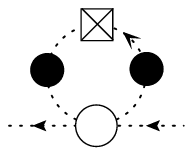}}
\caption{Flow equation of the ghost propagator, Eq. (\ref{full ghost flow formula})}
\label{ghost full flow}
\end{figure*}

The propagator flows are obtained from the flow equation for the 
effective action (\ref{F6-***}) by differentiating twice 
w.r.t.\ the fields. These derivations are detailed in Appendix 
\ref{app details deriv}, their outcome is represented diagrammatically 
in Figs. \ref{gluon full flow}, \ref{ghost full flow}. All propagators and vertices, denoted by 
black and white circles respectively, are fully dressed $k$-dependent 
correlation functions. This has to be compared with DSE equations for 
the propagators where all diagrams contain one bare vertex.  
\begin{figure*}[t] 
{\Large$k\partial_k$} \parbox{50pt}{\includegraphics[width=50pt,clip=]{gluon-propagator.eps}}{\Large$^{-1}=$}
{\Large$\;-\;$}\parbox{90pt}{\includegraphics[width=90pt,clip=]{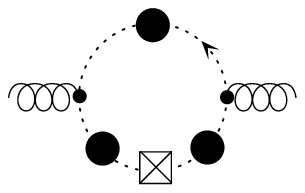}}
{\Large$\;-\;$}\parbox{90pt}{\includegraphics[width=90pt,clip=]{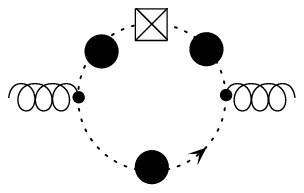}}
\caption{Truncated flow equation of the gluon propagator. Here and in the following, the bare vertices at $k=\Lambda$ are symbolized by small 
  dots.} 
\label{gluon trunc flow} 
\end{figure*} 
\begin{figure*}[t]  
{\Large$k\partial_k$} \parbox{50pt}{\includegraphics[width=50pt,clip=]{ghost-propagator.eps}}{\Large$^{-1}=$} 
\parbox{90pt}{\includegraphics[width=90pt,clip=]{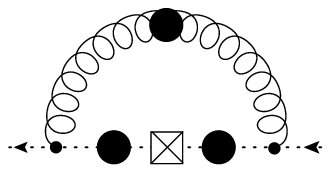}}
{\Large$\;+\;$}\parbox{90pt}{\includegraphics[width=90pt,clip=]{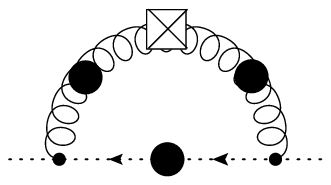}}
\caption{Truncated flow equation of the ghost propagator.} 
\label{ghost trunc flow}
\end{figure*} 

For a solution of these equations we have to approximate the full 
effective action. In the present work we approximate it by the classical action and fully momentum dependent (inverse) 
propagators.  We are specifically interested in the infrared where we 
assume scaling. Uniqueness of scaling as proven in the last 
section~\ref{sec:scaling} then implies ghost dominance. Consequently, 
we drop the gluonic vertices and only keep the ghost vertices. The 
resulting flow equations for the gluon and ghost propagators are shown 
in Figs. \ref{gluon trunc flow}, \ref{ghost trunc flow}. 
 
We pause here for a moment to discuss the meaning of and the 
justification for this truncation in detail. The generating functional 
(\ref{6-11}) is a functional integral defined by the full vacuum wave 
functional $\psi[A]$ which is, however, unknown. In Ref.\ \cite{RX1}, 
the vacuum functional has been determined explicitly to one-loop order 
through a perturbative solution of the Schr\"odinger equation for the 
Christ-Lee Hamiltonian. It was found that non-local terms in the 
couplings of two, three and four gluon fields in the wave functional 
give contributions to the static gluon propagator
that are relevant to its ultraviolet behavior, in particular 
its anomalous dimension. In higher loop orders, non-local terms in the 
couplings of more than four gluon fields will also become relevant to 
the gluon two-point function. By non-local we refer to coefficient 
functions that become singular for exceptional momenta. By neglecting 
these terms in the truncation considered in the present paper, the 
ultraviolet behavior of the two-point function will not be accurately 
reproduced, i.e., the power of the logarithmic correction in this 
kinematic regime will be incorrect. 
 
On the other hand, such non-local terms and three- and four-gluon 
couplings are not necessarily relevant to the infrared behavior, which 
is our main concern here: it has been argued \cite{Zwa92} that in the 
infrared the static ghost propagator is strongly enhanced 
relative to its tree level behavior, while the gluon propagator is 
suppressed or even vanishing. This is precisely what happens for the 
unique scaling solution as discussed in the previous section: the 
infrared behavior is dominated by those diagrams with the largest 
number of ghost propagators (``ghost dominance''). Indeed, the 
arguments about the kinematic singularities in \cite{Fischer:2009tn} 
can also be directly transferred to Coulomb gauge. They entail that 
for the scaling solution neither the non-local terms described  
before nor the couplings of three or more gluons in general contribute to  
those diagrams that dominate the infrared behavior. The same  
can be inferred from the diagrammatic analysis of 
Ref.\ \cite{RX1} when extended to higher perturbative orders, if 
one takes into account that there are no non-local terms or higher couplings 
including ghost fields in the ``action'' (\ref{6-14}).
 
The approximation of keeping a bare or tree-level ghost-gluon vertex 
is based on the ``non-renormalization theorem'' for this vertex 
\cite{Tay71}.  Although this theorem was originally formulated for QCD 
in Landau gauge and the space-time correlation functions, the argument 
carries over without change to the present situation. Indeed we have 
shown in the last section that it follows for the unique scaling 
solution. It has been confirmed on the non-perturbative level for the 
Landau gauge case in lattice studies \cite{CMM04}. As for the Coulomb 
gauge, a perturbative evaluation of the vertex (to one-loop level) at 
the symmetric point shows that the quantum corrections are finite and 
independent of the scale \cite{RX1,CRW09}. 
 
In summary, we can drop the gluonic vertices in the infrared without 
spoiling the quantitative nature of our approximation. We emphasize 
that for large momenta this is evidently not true. Finally, we also 
drop the tadpole diagrams in the flow equations for the static
propagators. The four-point couplings appearing in these 
diagrams are not contained in the integrand of the generating 
functional, but can build up in the course of the renormalization 
group flow. We simply assume that their contribution is negligible in 
the infrared, at least for the qualitative behavior of the two-point 
correlation functions, which leads us to the final form of the flow 
equations represented in Figs. \ref{gluon trunc flow}, \ref{ghost trunc flow}.

\section{Effective action and flows}  
 
\subsection{Parametrization of the effective action}\label{parameterisation}
 
The arguments above single out a specific approximation 
of the effective action. First of all, it relies on an expansion of 
the effective action in powers of the field, 
\begin{equation}\label{eq:effact}  
  \Gamma_k[\phi] =\sum_{N_A,N_c,N_{\bar c}} \0{1}{N_A!} \0{1}{N_c!}  
  \0{1}{N_{\bar c}!}\Gamma_{k,n_1\cdots n_N}^{(N)}  
  \cdot \phi_{n_1}\cdots \phi_{n_N}\,, 
\end{equation} 
where the $\phi_{n_i}$ stand for either the gluon fields ($\phi=A$) or 
the ghost fields ($\phi=c,\cb$).  In the following we take into 
account the bare ghost-gluon vertex and the full momentum 
dependent propagators. In an upgrade of the approximation we will also 
take into account tadpole terms related to ghost and ghost-gluon 
vertices.  This provides a first error estimate for the approximation 
scheme set up here. All other vertices are set to zero.  Therefore, in 
the minimal order of the approximation the only non-vanishing 
$n$-point functions are the ghost and gluon (inverse) propagators and 
the ghost-gluon vertex. 
 
We parametrize the inverse gluon propagator as follows,  
\begin{equation}\label{decomposition gluon} 
(2\pi)^{6} \fc{\dl^2\G_k}{\dl A^a_i(\fp)\dl A^b_j(\fq)} = \dl^{ab} 
t_{ij}(\fp) \, 2\om_k(p) (2\pi)^3\dl^3(\fp+\fq) \,.   
\end{equation} 
The diagonality in color space is due to global color symmetry. The 
transverse projector comes with the choice of Coulomb gauge where the 
gauge fields are transverse, see \eq{eq:transverse}, and momentum 
conservation arises from spatial translational invariance of the 
theory. The only quantity left to be determined by the flow equation 
is $\om_k(p)$, which depends only on the absolute value of the 
external momentum due to rotational invariance of the theory, and on 
the cutoff momentum $k$. The factor of $2$ is mere convention. In the 
flow we need the gluon propagator $G_{A,k} \, t_{ij}\delta^{ab}$ with the 
scalar function   
\begin{equation}\label{eq:GkA}  
G_{A,k}(p)=\0{1}{2 \omega_k(p)+R_{A,k}(p)}\,.  
\end{equation} 
The full gluon propagator at vanishing cut-off is given by 
$G_A(p)=1/2\omega(p)$, with $\omega(p)\equiv\omega_0(p)$.   
The ghost two-point function can be expressed as 
\begin{equation}\begin{split}\label{decomposition ghost} 
    -(2\pi)^{6} \fc{\dl^2\G_k}{\dl \cb^a(\fp)\dl c^b(\fq)}  
    =&\, \dl^{ab} g\fc{p^2}{d_k(p)} (2\pi)^3\dl^3(\fp+\fq) \;\;, 
\end{split}\end{equation} 
where $d_k(p)$ is the ghost form factor, which is the quantity to be 
calculated. The ghost propagator $G_{c,k} \, \delta^{ab}$ comprises the scalar  
function
\be\label{eq:barGkC}
G_{c,k}(p) =  \fc{1}{g}\bar{G}_{c,k}(p)
\ee
with
\begin{equation}\label{eq:GkC}  
\bar{G}_{c,k}(p)=\0{1}{p^2/d_k(p)+\bar{R}_{c,k}(p)}\,.  
\end{equation}
We have included an explicit constant factor of $1/g$ in the definition of the
ghost form factor for ease of comparison with the Dyson-Schwinger equations of the
variational approach in subsection \ref{analytical integration}.
The full ghost propagator at vanishing cut-off is  
$G_c(p)=d(p)/gp^2$, where $d(p)\equiv d_0(p)$.  
The last quantity to specify is the ghost-gluon vertex. We have argued 
in the previous section that it is well approximated by its bare part,  
\begin{equation}\begin{split}\label{eq:ghostgluon} 
 -(2\pi)^{9} \fc{\dl^3\G_k}{\dl \cb^a(\fp_1)\dl c^b(\fp_2)\dl A^c_i(\fp_3)} 
  =& \\  
 -igf^{abc}p_{1,j}t_{ij}(\fp_3) (2\pi)^3 & \dl^3(\fp_1+\fp_2+\fp_3)\,, 
\end{split}\end{equation}
where we have used the fact that the gauge field is transverse in  
Coulomb gauge. With these conventions, in
particular Eq. (\ref{decomposition ghost}), a suitable choice of the renormalization group invariant coupling $g_R$
is
\be\label{def coupling}
g_R(p) = d(p) (p/\omega(p))^{1/2} \;,
\ee
see \cite{Fischer:2005qe}.
\begin{figure*}[ht]
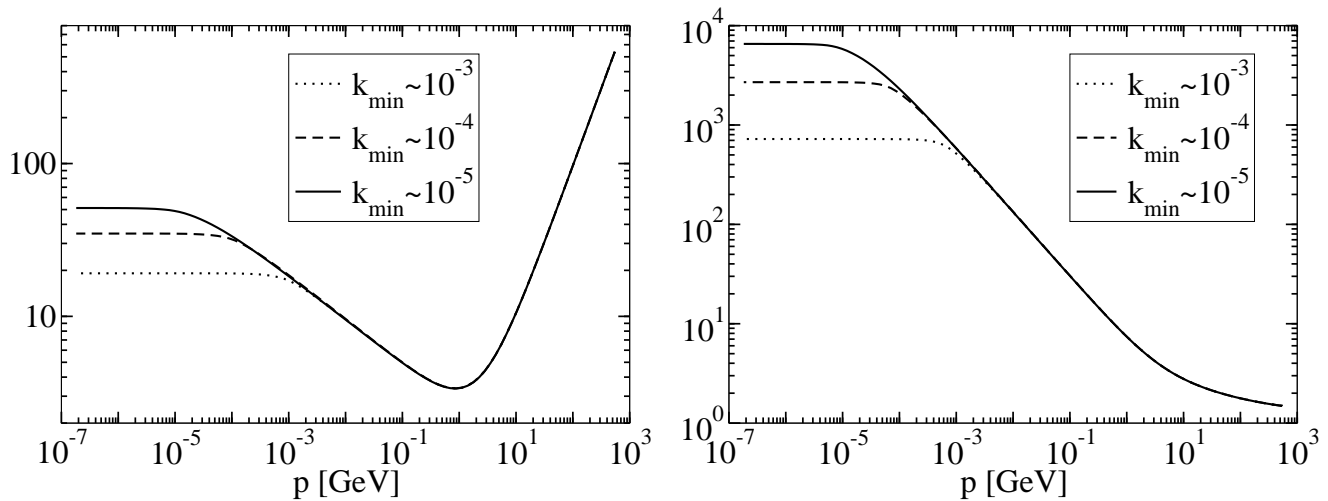
 
\hspace{\fill} 
\includegraphics[scale=0.35,clip=]{3diff_kmin_omega.eps} 
\hspace{\fill} 
\includegraphics[scale=0.35,clip=]{3diff_kmin_ghost.eps} 
\hspace{\fill} 
\caption{Gluon propagator $\om$ (left) and ghost dressing $d$ (right) for different 
  minimal cutoffs $k_{min}$} 
\label{diff kmin} 
\end{figure*}

\subsection{Approximation without tadpoles}\label{numerical integration} 
Now we are in the position to solve the flow equations for the 
propagators within the minimal truncation introduced above: the only 
non-vanishing vertex function is the bare ghost-gluon vertex 
\eq{eq:ghostgluon}. In particular this eliminates the tadpole 
diagrams. Inserting this approximation into the flow equations for the 
gluon and ghost propagators shown in Figs. \ref{gluon trunc flow}, \ref{ghost trunc flow} we arrive 
at 
\begin{widetext} 
\begin{eqnarray}\label{gluon flow final} 
  \partial_t\om_k(p) &=& - \fc{N_c}{2}\int \frac{d^3 q}{(2\pi)^3} \;  
 \left( \bar{G}_{c,k}\dot{\bar{R}}_{c,k} \bar{G}_{c,k}\right)(q)\, \bar{G}_{c,k}(|\fp+\fq|) 
  \;q^2(1-(\hat{\fp}\cdot\hat{\fq})^2) \,,\\[2ex]\nonumber  
    \partial_t  d_k^{-1}(p) &= &N_c  p^2 \int \frac{d^3 q}{(2\pi)^3} \Bigl[ 
\left( G_{A,k} \dot{R}_{A,k} G_{A,k}\right)(q)\, 
    \bar{G}_{c,k}(|\fp+\fq|) \\[1ex]  
&& \hspace{2cm}+ \left( \bar{G}_{c,k}\dot{\bar{R}}_{c,k} \bar{G}_{c,k}\right)(q)\, G_{A,k}(|\fp+\fq|) \;  
\0{q^2}{(\fp+\fq)^2}  \Bigr]  (1-(\hat{\fp}\cdot\hat{\fq})^2) \,. 
 \label{ghost flow final} \end{eqnarray} 
\end{widetext} 
The computation is detailed in Appendix C. Eqs. (\ref{gluon flow final}) and 
(\ref{ghost flow final}) are two coupled functional ordinary differential 
equations for $\omega_k$ and $d_k$, which can be solved numerically. 
Due to our definition (\ref{decomposition ghost}) of the ghost form
factor $d_k(p)$ the bare coupling constant $g$ has formally disappeared from the
propagator flow equations. Let us stress again that a physically meaningful
renormalization group invariant coupling is defined by Eq. (\ref{def coupling}). 
To incorporate the appropriate initial conditions 
it is convenient to cast the differential flow equations into an 
integral form,
\begin{equation}\label{int flow om symbolical} 
    \om_k(p) - \om_\Lambda(p) = \int^k_\Lambda\frac{dk'}{k'} 
    \int\frac{d^3 \ell}{(2\pi)^3} \; I^\omega_{k'}[d_{k'}](\fl,\fp) \;, \\ 
\end{equation}
\begin{equation}\label{int flow d symbolical} 
    d_k^{-1}(p) - d_\Lambda^{-1}(p) = \int^k_\Lambda \frac{dk'}{k'} 
    \int\frac{d^3 \ell}{(2\pi)^3} 
    \; I^d_{k'} [\om_{k'},d_{k'}](\fl ,\fp)\;. 
\end{equation} 
$I^\omega$ and $I^d$ stand for the integrands of the loop integrals on 
the r.h.s. of Eqs. (\ref{gluon flow final}) and (\ref{ghost flow 
  final}).  
 
The initial conditions $d_\Lambda(p),\omega_\Lambda(p)$ for the flow
can be determined by perturbation theory \cite{RX1,CRW09}. Due to the
kinematic structure of the ghost-gluon vertex no mass term for the
ghost is produced. Moreover, contributions with higher powers of
momenta than the ones in lowest-order perturbation theory, $d^{(0)}
(p) = 1$ and $\omega^{(0)} (p) = p$, are suppressed by the
corresponding powers of $\Lambda$.  Note, however, that there is 
additional logarithmic scaling.  In the case of
$d_\Lambda(p)$, this implies that for a large initial cut-off scale
$k=\Lambda$ we only have to fix the constant $d_\Lambda (p) \equiv
d_\Lambda$.

\begin{figure}[h]  
\includegraphics[scale=0.8,clip=]{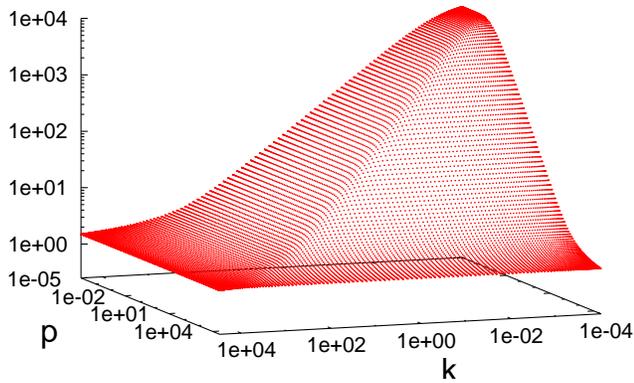} 
\caption{The flow of the full ghost form factor, $d_k(p)$, is 
  shown. The gradual formation of the IR power law when lowering the 
  cutoff scale $k$ is explicitly seen.} 
\label{ghost full flow 3dim pic} 
\end{figure}

For the gluon the introduction of the regulator term enforces a  
mass-like term, i.e., a $p$-independent contribution to  
$\omega_k (p)$,
due to the modified Slavnov-Taylor identities (mSTI), see e.g.\ \cite{ 
  Ellwanger:1994iz,Ellwanger:1995qf,Freire:2000bq,Pawlowski:2005xe,Fischer:2008uz}.  We 
rush to add that the direct use of the mSTI in the infrared is not 
advisable as the mSTI only fix the 
longitudinal mass which does not relate to the transverse one in the 
infrared \cite{Pawlowski,Fischer:2008uz,Pawlowski:2005xe}. We 
conclude that for large initial cut-off scales $k=\Lambda$ the inverse 
gluon propagator has two relevant parameters, the mass-like parameter  
$a$ and the coefficient of the classical term.  The latter is put to one and 
we have $\omega_\Lambda(p) = p+a$.  
 
It is evident from the form of the flow equations in \eq{int flow om symbolical}, \eq{int flow d symbolical} that the solution will not show infrared scaling 
unless the parameter $d_\Lambda^{-1} (p) \equiv d_\Lambda^{-1}$ is 
fine-tuned (at least for $\beta = 2\kappa_c > 0$). Such 
fine-tuning of relevant parameters is a well-known 
initial condition problem for RG flows. Indeed, it also occurs in 
Landau gauge Yang-Mills theory, see e.g. 
\cite{Pawlowski,Fischer:2008uz,Pawlowski:2003hq}, where 
it is directly related to the resolution of the Gribov problem, see 
\cite{Fischer:2008uz,vonSmekal:2008ws}. In \cite{Fischer:2008uz} it 
has also been shown that there is a full one parameter family of 
solutions compatible with non-perturbative renormalization where only 
the endpoint shows a scaling behavior whereas the other solutions 
show a decoupling behavior: a gluon with a mass-like propagator and a 
ghost which is at most logarithmically enhanced. Such a scenario very 
likely also applies to Coulomb gauge and hence deserves further 
investigation. 
 
In the present case we have numerically solved the fine-tuning 
condition for $d_\Lambda$ with the constraint of infrared scaling  
for the ghost dressing function. The parameter $a$ in the initial 
condition $\omega_\Lambda(p)$ is fixed by demanding that $\omega(p)$ reduce 
to the perturbative form $\omega(p) \propto p$ for ``large'' momenta $p$  
close to but below $\Lambda$. The regulator used 
in the numerical solution is  
\begin{equation}\label{explicit shape} 
r_k(p) =  \exp\lf(\fc{k^2}{p^2} -\fc{p^2}{k^2}\ri) \;\;. 
\end{equation} 
Our numerical procedure is detailed in Appendix~\ref{numerics}. The 
results for the inverse gluon propagator $\omega_k(p)$ and the ghost 
dressing function $d_k(p)$ are shown in Fig.~\ref{diff kmin} for 
different values of the minimal cutoff $k_{min}$ down to which the 
flow integration has been carried out. It is seen that the power law 
behavior in both cases extends towards the IR as the cutoff $k_{min}$ 
is lowered, although we have implemented a scaling behavior - not the 
``horizon condition'' $d_0^{-1} (p=0) = 0$ - only  
for the ghost 
dressing. Thus, we arrive at the nontrivial result that a solution for 
the flow equations can be found that obeys a power law behavior 
\eq{eq:scalpropsCoul} for both the gluon and the ghost propagator.  For the 
sake of illustration, we also display the full flow of the ghost 
dressing function, $d_k(p)$, in Fig.~\ref{ghost full flow 3dim pic}. 
 
\begin{figure*}[ht]
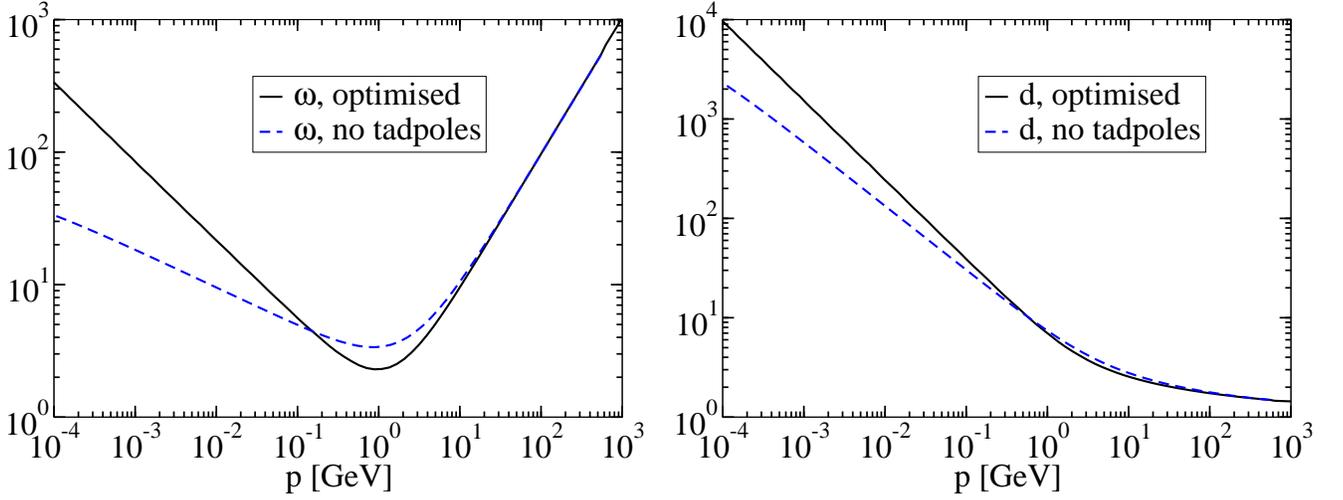
 
\hspace{\fill} 
\includegraphics[scale=0.35,clip=]{compare_omega_approx_full.eps} 
\hspace{\fill} 
\includegraphics[scale=0.35,clip=]{compare_ghost_approx_full.eps} 
\hspace{\fill} 
\caption{Comparison of the gluon propagators $\om$ (left) and the ghost 
  dressings $d$ (right) calculated from the optimized flow and from the flow without tadpoles.} 
\label{comp opt to no tads} 
\end{figure*}

The IR power laws are extracted from the numerical solution shown in 
Fig. \ref{diff kmin}. The scaling coefficients $\alpha,\beta$ defined in  
\eq{eq:scalpropsCoul} are computed as  
\begin{equation}\label{def alpha beta} 
\omega(p\to 0) \sim p^{- \alpha}\,, \qquad d(p\to 0) \sim p^{- \beta} \, . 
\end{equation} 
and with the numerical solution we get  
\begin{equation}\label{IR powers full} 
 \alpha=0.28,\qquad \beta=0.64\;,  
\end{equation} 
and hence $\alpha$ and $\beta$ satisfy the sum rule \eq{eq:Cscaling}, 
already found for the Coulomb gauge DSE in 
\cite{Schleifenbaum:2006bq}.  Note however that the scaling 
coefficients obtained in the present truncation differ from the ones 
obtained in the DSE. In Fig. \ref{comp opt to no tads} the solutions of the 
FRG flow equation for $\omega (p)$ and $d(p)$ are 
compared to the results obtained from an optimized calculation in sect. \ref{analytical integration} which in turn are precisely the results found in \cite{Feuchter:2004mk} by a 
variational calculation, see Fig. \ref{flow simplif}. This variational calculation gave rise to the DSEs which will be found in 
sect. \ref{analytical integration} as approximation to the full flow 
equation. While the curves in Fig. \ref{comp opt to no tads} match in the UV, the results of the FRG in 
the present minimal truncation are less infrared enhanced than the 
ones of the DSE. Note that the scaling coefficients are 
expected to depend on the 
chosen regulator. It has been already proven in 
\cite{Pawlowski:2003hq} for Landau gauge Yang-Mills theory that the scaling 
coefficients of FRG and DSE agree for optimized regulators if a 
bare ghost-gluon vertex is used. Optimized regulators are those 
that maximize the physics content of a given truncation scheme at a 
given order, for details of the optimization theory for the FRG we 
refer the reader to 
\cite{Litim:2000ci,Pawlowski:2005xe}. The 
details of the arguments put forward in \cite{Pawlowski:2003hq} 
directly carry over to the Coulomb gauge system.

\subsection{Optimization}\label{analytical integration}

In this chapter we shall use the optimization arguments hinted at 
above in order to optimize the physics content of our present 
truncation. Similar, but more refined arguments have been used in 
Landau gauge for arriving at FRG results for the propagators that 
quantitatively agree with the lattice results 
\cite{Pawlowski,Fischer:2008uz}, the details will be published in 
\cite{Pawlowski}. Here we first follow the arguments put forward in 
\cite{Pawlowski:2003hq}. To that end consider the following 
approximation: under the loop integrals we replace the propagators at 
the running momentum scale $k, \omega_k$ and $d_k$, by the propagators 
of the full theory, i.e. the ones at zero scale $k = 0$: 
\begin{equation} 
\label{11-41} 
d_k(p) \rightarrow d_{k = 0}(p)  \; ,\;  
\om_k(p) \rightarrow \om_{k = 0}(p) \;.
\end{equation} 
Note that this does only imply that the difference between the 
propagators at $k=0$ and the regularized ones at $k$ drops out in the 
integrals. Indeed one can explicitly construct 
regulators for which this holds true in the asymptotic IR region, see 
\cite{Pawlowski:2003hq}. Note that due to the strong infrared 
suppression introduced with the regulator choice \eq{explicit shape} 
the approximation \eq{11-41} is quantitatively reliable inside
the loop integrals except for a small range of momenta $p$ around the
scale $k$. The approximation \eq{11-41} allows us 
to analytically integrate the flow equations (\ref{gluon flow final}), 
(\ref{ghost flow final}) over $k$. The only $k$-dependence left is the 
explicit one on the regulator as the vertices are not $k$-dependent 
from the outset. Hence the flow can be rewritten as a total 
$t$-derivative of the loop integrals with full propagators and we 
arrive at 
\begin{widetext} 
\begin{eqnarray}\label{ghost simple} 
  (d_0^{-1} - d_\Lambda^{-1})(p) &= &\lf. -N_c  \int  
\frac{d^3 q}{(2\pi)^3} 
  \0{1}{2\om_0(q)+R_{A,k}(q)} \0{(1-(\hat{\fp}\cdot\hat{\fq})^2) }{ 
    (\fp+\fq)^2 d^{-1}_0(|\fp+\fq|)+\bar{R}_{c,k}(| 
    \fp+\fq|)} \ri|^{k=0}_{k=\Lambda} \,,\\[1ex] 
\label{gluon simple} 
(\om_0 - \om_\Lambda)(p) &= &\lf. \fc{N_c}{4} \int  
\frac{d^3 q}{(2\pi)^3} \; \0{q^2}{q^2 d^{-1}_0(q)+\bar{R}_{c,k}(q) }  
\0{(1-(\hat{\fp}\cdot\hat{\fq})^2) }{ (\fp+\fq)^2 
  d^{-1}_0(|\fp+\fq|)+\bar{R}_{c,k}(|\fp+ 
  \fq|)} \ri|^{k=0}_{k=\Lambda} \;\;.
\end{eqnarray} 
\end{widetext}
We notice that the flow equations \eq{ghost 
  simple}, \eq{gluon simple} have acquired the form of DSEs. Given the 
fact that $R_{k=0}=0$ (see Eq. (\ref{regulator properties}) these equations coincide with the DSEs obtained 
in \cite{Feuchter:2004mk} (to be precise, the DSE for $\omega (p)$ in
\cite{Feuchter:2004mk} contains additional contributions which are, however,
subleading in the infrared), with a different UV-regularization 
realized here through the $(k = \Lambda)$-terms.  
Moreover, the optimization arguments in 
\cite{Pawlowski:2003hq,Pawlowski:2005xe} imply that the flows 
\eq{ghost simple}, \eq{gluon simple} provide the best approximation to 
the full theory for the IR asymptotics. In \cite{Pawlowski} it is 
shown that in the given truncation this argument extends to the full 
momentum regime: by adding the tadpole diagrams related to ghost-only 
and ghost-gluon vertices to the flow equations displayed in 
Figs. \ref{gluon trunc flow}, \ref{ghost trunc flow} and using the DSE for the tadpole vertices in 
the flows one can show that this leads to the integrated flow 
\eq{ghost simple}, \eq{gluon simple}. If we also add the gluonic 
diagrams, this argument gets more involved. 
 
It remains to adjust the initial conditions. We could proceed in 
the same way as for the the numerical solution in the last subsection to 
implement the condition of infrared scaling for the ghost dressing 
function. However, it is much simpler to use as an input the information  
from this numerical solution that $\beta = 2 \kappa >0$ and thus 
$d_0^{-1} (p=0) = 0$
(the horizon condition), so we can write 
\begin{equation}\label{ghost simple subtracted}
  d_0^{-1}(p) = \lf.\int \frac{d^3 q}{(2\pi)^3} \,[\mathrm{int}(k,\fp,\fq)  
  - \mathrm{int}(k,\fp=0,\fq)]\ri|^{k=0}_{k=\Lambda} \;\;, 
\end{equation} 
where $\mbox{int}(k, \fp, \fq)$ denotes the integrand in (\ref{ghost 
  simple}). More details of the numerical procedure can be found in  
Appendix~\ref{numerics}. 
 
\begin{figure}  
\includegraphics[scale=0.35,clip=]{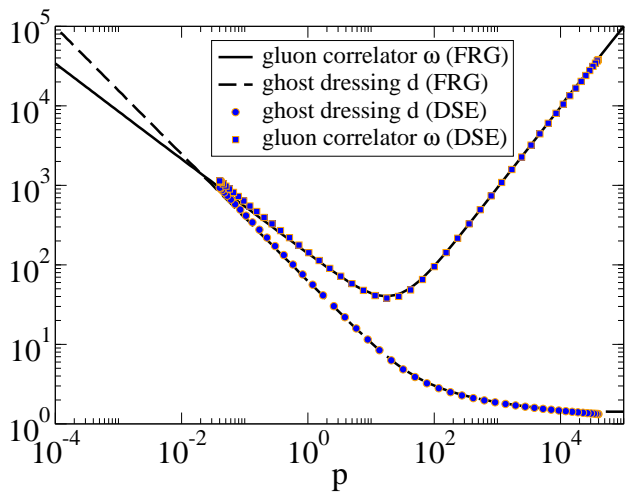} 
\caption{Gluon propagator $\om$ and ghost dressing function $d$ 
  from the optimized flow equation in comparison with the results of 
  the DSEs obtained from the variational ansatz in 
  \cite{Feuchter:2004mk}. The results lie on top of each other as 
  expected.} 
\label{flow simplif} 
\end{figure}

The results of the iteration are shown in Fig. \ref{flow simplif}. 
A power law as in Eq. (\ref{def alpha beta}) emerges in the infrared region for both the gluon 
energy $\om(p)$ and the ghost dressing function $d(p)$ with the IR exponents 
\begin{equation}\label{IR powers DSE} 
\alpha=0.60,\qquad \beta=0.80\;, 
\end{equation} 
which is precisely one of the two possible IR solutions found
analytically for the DSE in \cite{Schleifenbaum:2006bq}. Furthermore,
it corresponds to the solution found in the variational approach
\cite{Feuchter:2004mk}. It must be noted that the second possible
solution in the analytical approach of Ref.\
\cite{Schleifenbaum:2006bq} has also been confirmed in a numerical
variational calculation \cite{Epple:2006hv}. Indeed, it is also
present in our optimized approximation which has the identical
infrared properties as the DSE. However, it is not clear to us at
present whether after the inclusion of the gluonic diagrams this
solution persists as an infrared stable one. Note in this context that
it requires additional fine-tuning and hence may be unstable.

\begin{figure}  
\includegraphics[scale=0.35,clip=]{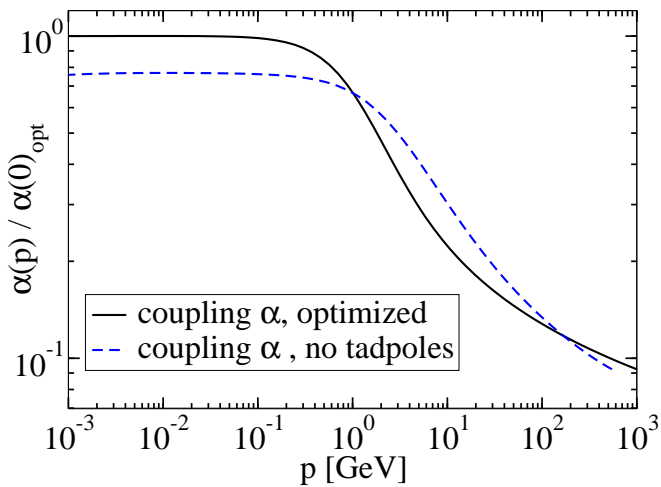} 
\caption{The running coupling constant $\alpha = g_R^2/4\pi$, Eq. (\ref{def coupling}), calculated from the optimized flow and from the flow without tadpoles.} 
\label{running coupling} 
\end{figure} 

Finally, in Fig. \ref{running coupling}, we show the running coupling constant $\alpha = g_R^2/4\pi$, see Eq. (\ref{def coupling}), calculated by using the propagators from the optimized flow equation as well as from the flow equation without the tadpoles. The plateau in the IR is due to the sum rule (\ref{eq:Cscaling}) which is fulfilled by the propagators resulting from both approximations of the flow equations.

\section{Conclusions} 

We have presented a new approach to the non-perturbative calculation
of static propagators, or correlation functions at equal times, from
the Hamiltonian formulation of a quantum field theory. In the
generating functional of the correlation functions at equal times, (minus twice) the logarithm
of the modulus of the vacuum wave functional comes to play the role of the Euclidean
action in the usual covariant theory.  We have then adapted the functional
renormalization group to this Hamiltonian formulation.

This new approach has subsequently been applied to Yang-Mills theory in the
Coulomb gauge. The derivation of the corresponding flow equations for
the propagators has been presented in detail. In order to arrive at a closed
system of equations, we have replaced the dressed ghost-gluon vertex with
the corresponding bare one (or a vertex with constant dressing, a good
approximation according to perturbative arguments and lattice
calculations), and we have neglected all tadpole diagrams 
and vertices with three or more gluon lines. We have also presented an
approximation that allows for an analytical integration of the flow
equations and have argued that it actually corresponds to an optimized
choice of the regulator functions.  We have discussed in detail the choice
of the initial conditions which serve to implement the normalization
conditions corresponding to an infrared scaling solution for the
propagators.

The result of the numerical solution of the flow equations has been compared
to the solution of a system of Dyson-Schwinger equations derived from
a variational principle for the vacuum wave functional. The solution
of the flow equations agrees with one of the two possible scaling
solutions in the latter approach. In the approximation without
tadpoles we have obtained slightly different values for the infrared anomalous
dimensions than from the Dyson-Schwinger equations. Indeed, in this
truncation the anomalous dimensions mildly depend on the regulator
functions and only agree with the Dyson-Schwinger values for optimized
regulators, as has been argued in \cite{Pawlowski:2003hq}. In the
optimized approximation the infrared anomalous dimensions are
regulator-independent. This supports the reliability of the optimized
approximation. 

We have not presented the second scaling solution which poses an
additional fine-tuning problem. Its resolution also allows to study
the interesting question of infrared stability of this solution, and
will be discussed elsewhere. 

Note also that the formalism put forward in the present work allows to
directly access the confining properties of the propagators. This can be done by using the Wilson loop potential evaluated in \cite{Braun:2007bx}.  Moreover
one can study full dynamical QCD in Coulomb gauge along the lines of
\cite{Braun:2009gm}. A first step in this direction is to implement
the missing gluonic diagrams for a comparison with lattice data in the
full momentum regime. These investigations will certainly shed more
light on the pressing unresolved questions of low energy QCD.

\begin{acknowledgments}
M.L. was supported by the Internationales Graduiertenkolleg ``Hadronen im Vakuum, in Kernen und Sternen''.
J.M.P. acknowledges support by Helmholtz Alliance HA216/EMMI.
H.R. acknowledges support by DFG-Re856/6-3.
A.W. would like to thank the Deutscher Akademischer Austauschdienst (DAAD), Conacyt 
project no.\ 46513-F, and CIC-UMSNH for financial support.
\end{acknowledgments} 
 
\appendix 
 
\section{\protect\label{app details deriv act}Details of the
  derivation of the flow of the effective action}
 
In this appendix we fill in some of the details of the derivation of the flow equation (\ref{F6-***}). 
From the regularized generating functional (\ref{5-13}) and  
Eq.\ (\ref{regulator term}), we derive the flow equation 
\begin{equation} 
\partial_t Z_k[J,\s,\sb]  
= (-\lfc{1}{2} \lfc{\dl}{\dl J}\cdot \dot{R}_{A,k}\cdot \lfc{\dl}{\dl J} + \lfc{\dl}{\dl\s}\cdot\dot{R}_{c,k}\cdot \lfc{\dl}{\dl \sb}) Z_k[J,\s,\sb] \;, 
\end{equation} 
where the difference in signs of the gluon and ghost regulator term compared to Eq. (\ref{regulator term}) is due to the Grassmann property of the ghost fields and sources. 
 
The definition of the Schwinger functional $W_k$ generating connected Green functions reads 
\begin{equation} 
W_k[J,\s,\sb] := \ln Z_k[J,\s,\sb] \;,
\end{equation} 
and therefore its flow is 
\begin{equation}\begin{split}\label{flow of W} 
\partial_t  W_k = &\lf(-\fc{1}{2} \fc{\dl W_k}{\dl J}\cdot \dot{R}_{A,k}\cdot \fc{\dl W_k}{\dl J} - \fc{1}{2} \Tr\; \dot{R}_{A,k} \fc{\dl^2 W_k}{\dl J\dl J}\ri. \\  
&+ \lf.\fc{\dl W_k}{\dl\s}\cdot\dot{R}_{c,k}\cdot \fc{\dl W_k}{\dl \sb} -\Tr\; \dot{R}_{c,k} \fc{\dl^2 W_k}{\dl\sb\dl\s}\ri) \;. 
\end{split}\end{equation} 
The effective action $\G_k$ is defined via a modified Legendre transformation, 
\begin{equation}\begin{split}\label{Legendre transformation} 
\G_k[A,\cb,c] := &-W_k[J_k,\s_k,\sb_k] + J_k\cdot A + \sb_k\cdot c + \cb\cdot \s_k \\ 
&- \fc{1}{2} A\cdot R_{A,k}\cdot A - \cb\cdot R_{c,k}\cdot c \;, 
\end{split} 
\end{equation} 
which turns into the usual one upon taking $k\rightarrow 0$, because $R_{A,k=0} = R_{c,k=0} = 0$. In Eq. (\ref{Legendre transformation}), the sources are functionals of the fields (whose notation is suppressed), which are the expectation values of the corresponding field operators. The relations between sources and fields are given by
\begin{equation}\begin{split} 
A_i^a(-\fp) =& \,(2\pi)^3\fc{\dl W_k[J_k,\s_k,\sb_k]}{\dl J_i^a(\fp)} \;,\\ 
c^a(-\fp) =& \,(2\pi)^3\fc{\dl W_k[J_k,\s_k,\sb_k]}{\dl \sb^a(\fp)} \;, \\ 
\cb^a(-\fp) =& -(2\pi)^3\fc{\dl W_k[J_k,\s_k,\sb_k]}{\dl \s^a(\fp)} \;, 
\end{split}\end{equation} 
or shorthand 
\begin{equation} 
A^a = \fc{\dl W_k}{\dl J^a} \,,\qquad c = \fc{\dl W_k}{\dl\sb} \,,\qquad \cb = -\fc{\dl W_k}{\dl\s} \;. 
\end{equation} 
With these definitions and with Eq. (\ref{flow of W}), the flow of the effective action is written as 
\begin{equation}\begin{split}\label{flow of G with W} 
\partial_t\G_k =& -(\partial_t W_k)[J_k,\s_k,\sb_k] \\ 
&- \partial_t J_k\cdot \fc{\dl W_k}{\dl J} - \partial_t\s_k\cdot\fc{\dl W_k}{\dl \s} - \partial_t\sb_k\cdot\fc{\dl W_k}{\dl\sb} \\ 
&+ A\cdot\partial_t J_k + \partial_t\sb_k\cdot c - \partial_t \s_k\cdot\cb \\ 
&- \lfc{1}{2}A\cdot\dot{R}_{A,k}\cdot A - \cb\cdot\dot{R}_{c,k}\cdot c \\ 
=& \lf.\lf(\fc{1}{2} \Tr\; \dot{R}_{A,k} \fc{\dl^2 W_k}{\dl J\dl J} +\Tr\; \dot{R}_{c,k} \fc{\dl^2 W_k}{\dl\sb\dl\s}\ri)\ri|_{J_k,\s_k,\sb_k}\def\Tr{\mbox{Tr}} \;\;. 
\end{split}\end{equation} 
When expressed in the superfield notation introduced from Eq. (\ref{superfields}) on, this equation turns into Eq. (\ref{flow gamma with W}) and finally into Eq. (\ref{F6-***}). In components of the superfield, Eq. (\ref{F6-***}) reads 
\begin{widetext} 
\begin{equation}\label{flow of action} 
\partial_t\G_k = \fc{1}{2}\, \Tr\;  
\begin{pmatrix}[2.5] 
\dot{R}_{A,k}&&\\ 
&-\dot{R}_{c,k}&\\ 
&&-\dot{R}_{c,k}^T\\ 
\end{pmatrix} 
\lf.\begin{pmatrix}[2.5] 
\df{\dl^2 \G_k}{\dl A\dl A}+R_{A,k} & \df{\dl^2 \G_k}{\dl A\dl c} & \df{\dl^2 \G_k}{\dl A\dl \cb}\\ 
-\df{\dl^2 \G_k}{\dl \cb\dl A} & -\df{\dl^2 \G_k}{\dl \cb\dl c}+R_{c,k} & -\df{\dl^2 \G_k}{\dl \cb\dl \cb}\\ 
\df{\dl^2 \G_k}{\dl c\dl A} & \df{\dl^2 \G_k}{\dl c\dl c} & \df{\dl^2 \G_k}{\dl c\dl \cb}+R_{c,k}^T\\ 
\end{pmatrix}\ri.^{-1} \;\;. 
\end{equation} 
\end{widetext} 
From this equation we derive the flow equations for the ghost and gluon fields by taking functional derivatives w.r.t. these fields.

\section{Details of the derivation of the propagator flows}\label{app details deriv} 
 
In this section we derive the flow equations for the propagators from 
the flow equation for the effective action, Eq. (\ref{flow of 
  action}).  A useful relation for the following concerns the 
commutation of
fermionic derivatives past supermatrices. Consider a matrix 
with a block structure of commuting ($c$) and anticommuting ($a$) 
quantities and an anticommuting $\eta$: 
\begin{equation} 
  \eta \begin{pmatrix}c&a&a\\a&c&c\\a&c&c\end{pmatrix} = \begin{pmatrix}c&-a&-a\\-a&c&c\\-a&c&c\end{pmatrix}\eta  
  = M\begin{pmatrix}c&a&a\\a&c&c\\a&c&c\end{pmatrix}M\;\eta \;. 
\end{equation} 
This matrix structure is shared by $\dl^2\G_k/\dl\bar{\phi}\dl\phi$ as 
well as $\lf(\dl^2\G_k/\dl\bar{\phi}\dl\phi + \mathcal{R}_k\ri)^{-1}$, 
the latter because of Eq. (\ref{inversion relation}). 
 
In the following, $i$ and $j$ are condensed external indices. They 
stand for color indices, momenta and, in the case of gluon fields, 
also vector indices at the same time. They are, however, not part of 
the matrix notation. From 
\begin{widetext} 
\begin{equation}\begin{split} 
0 &= \fc{\dl}{\dl c_i}\lf(\lf(\fc{\dl^2\G_k}{\dl\bar{\phi}\dl\phi}+ \mathcal{R}_k\ri)\lf(\fc{\dl^2\G_k}{\dl\bar{\phi}\dl\phi}+ \mathcal{R}_k\ri)^{-1}\ri) \\ 
&= \fc{\dl^3\G_k}{\dl c_i\dl\bar{\phi}\dl\phi}\lf(\fc{\dl^2\G_k}{\dl\bar{\phi}\dl\phi} + \mathcal{R}_k\ri)^{-1}  
+ M \lf(\fc{\dl^2\G_k}{\dl\bar{\phi}\dl\phi} + \mathcal{R}_k\ri) M \fc{\dl}{\dl c_i} \lf(\fc{\dl^2\G_k}{\dl\bar{\phi}\dl\phi} + \mathcal{R}_k\ri)^{-1} 
\end{split}\end{equation} 
it follows that 
\begin{equation} 
\fc{\dl}{\dl c_i} \lf(\fc{\dl^2\G_k}{\dl\bar{\phi}\dl\phi} + \mathcal{R}_k\ri)^{-1} 
= - M \lf(\fc{\dl^2\G_k}{\dl\bar{\phi}\dl\phi} + \mathcal{R}_k\ri)^{-1} M \fc{\dl^3\G_k}{\dl c_i\dl\bar{\phi}\dl\phi} 
\lf(\fc{\dl^2\G_k}{\dl\bar{\phi}\dl\phi} + \mathcal{R}_k\ri)^{-1} \;\;. 
\end{equation} 
For bosonic derivatives the same formula holds without the $M$'s. 
Using this, we can derive the ghost propagator flow equation from the flow of the effective action, Eq. (\ref{flow of action}): 
\begin{equation}\label{derivation ghost flow}\begin{split} 
    \fc{\dl^2\dot{\G}_k}{\dl\cb_j\dl c_i} = &\fc{1}{2}\fc{\dl^2}{\dl\cb_j\dl c_i}\STr\lf[ \dot{\mathcal{R}}_k \lf(\fc{\dl^2\G_k}{\dl\bar{\phi}\dl\phi} + \mathcal{R}_k\ri)^{-1} \ri] \\ 
    = &\fc{1}{2}\fc{\dl}{\dl\cb_j} \STr \lf[- \dot{\mathcal{R}}_k \;M\lf(\fc{\dl^2\G_k}{\dl\bar{\phi}\dl\phi} + \mathcal{R}_k\ri)^{-1}M \fc{\dl^3\G_k}{\dl c_i\dl\bar{\phi}\dl\phi} \lf(\fc{\dl^2\G_k}{\dl\bar{\phi}\dl\phi} + \mathcal{R}_k\ri)^{-1} \ri] \\ 
    = &\fc{1}{2}\STr \lf[\dot{\mathcal{R}}_k \lf(\fc{\dl^2\G_k}{\dl\bar{\phi}\dl\phi} + \mathcal{R}_k\ri)^{-1} M \fc{\dl^3\G_k}{\dl\cb_j\dl\bar{\phi}\dl\phi} \lf(\fc{\dl^2\G_k}{\dl\bar{\phi}\dl\phi} + \mathcal{R}_k\ri)^{-1} M \fc{\dl^3\G_k}{\dl c_i\dl\bar{\phi}\dl\phi} \lf(\fc{\dl^2\G_k}{\dl\bar{\phi}\dl\phi} + \mathcal{R}_k\ri)^{-1} \ri] \\ 
    &-\fc{1}{2}\STr \lf[\dot{\mathcal{R}}_k \lf(\fc{\dl^2\G_k}{\dl\bar{\phi}\dl\phi} + \mathcal{R}_k\ri)^{-1} M \fc{\dl^3\G_k}{\dl c_i\dl\bar{\phi}\dl\phi} \lf(\fc{\dl^2\G_k}{\dl\bar{\phi}\dl\phi} + \mathcal{R}_k\ri)^{-1} M \fc{\dl^3\G_k}{\dl\cb_j\dl\bar{\phi}\dl\phi} \lf(\fc{\dl^2\G_k}{\dl\bar{\phi}\dl\phi} + \mathcal{R}_k\ri)^{-1} \ri] \\ 
    &-\fc{1}{2}\STr \lf[\dot{\mathcal{R}}_k 
    \lf(\fc{\dl^2\G_k}{\dl\bar{\phi}\dl\phi} + \mathcal{R}_k\ri)^{-1} 
    \fc{\dl^4\G_k}{\dl\cb_j\dl c_i\dl\bar{\phi}\dl\phi} 
    \lf(\fc{\dl^2\G_k}{\dl\bar{\phi}\dl\phi} + \mathcal{R}_k\ri)^{-1} 
    \ri] \;\;. 
\end{split}\end{equation} 
Setting the fields to zero, $A=\cb=c=0$, only the block matrices with the same number of ghosts and antighosts remain. With the definition 
\begin{equation} 
\mathcal{G}_k := \begin{pmatrix}[2.0]\lf(\df{\dl^2\G_k}{\dl A\dl A}+R_{A,k}\ri)^{-1}&0&0\\0&\lf(-\df{\dl^2\G_k}{\dl\cb\dl c}+R_{c,k}\ri)^{-1}&0\\0&0&\lf(\df{\dl^2\G_k}{\dl c\dl\cb}+R_{c,k}^T\ri)^{-1}\\\end{pmatrix} 
\end{equation} 
the first of the three terms in Eq. (\ref{derivation ghost flow}) becomes 
\begin{equation} 
  \fc{1}{2}\,\Tr \begin{pmatrix}[2.0]\dot{R}_{A,k}&0&0\\0&-\dot{R}_{c,k}&0\\0&0&- 
    \dot{R}_{c,k}^T\\\end{pmatrix} \mathcal{G}_k 
  \begin{pmatrix}[2]0&\df{\dl^3\G_k}{\dl\cb_j\dl A\dl c}&0\\0&0&0\\ 
    -\df{\dl^3\G_k}{\dl\cb_j\dl c\dl A}&0&0\\\end{pmatrix} 
  \mathcal{G}_k 
  \begin{pmatrix}[2]0&0&\df{\dl^3\G_k}{\dl c_i\dl A\dl\cb}\\ 
    \df{\dl^3\G_k}{\dl c_i\dl\cb\dl A}&0&0\\0&0&0\\\end{pmatrix} 
  \mathcal{G}_k 
\end{equation} 
\begin{equation}\begin{split} 
    =&\fc{1}{2}\,\Tr \;\dot{R}_{A,k} \lf(\df{\dl^2\G_k}{\dl A\dl 
      A}+R_{A,k}\ri)^{-1} \df{\dl^3\G_k}{\dl\cb_j\dl A\dl c} 
    \lf(-\df{\dl^2\G_k}{\dl\cb\dl c}+ R_{c,k}\ri)^{-1} 
    \df{\dl^3\G_k}{\dl c_i\dl\cb\dl A} 
    \lf(\df{\dl^2\G_k}{\dl A\dl A}+R_{A,k}\ri)^{-1} \\ 
    &+\fc{1}{2}\,\Tr \;\dot{R}_{c,k}^T \lf(\df{\dl^2\G_k}{\dl 
      c\dl\cb}+R_{c,k}^T\ri)^{-1} \df{\dl^3\G_k}{\dl\cb_j\dl c\dl A} 
    \lf(\df{\dl^2\G_k}{\dl A\dl A}+R_{A,k}\ri)^{-1} \df{\dl^3\G_k}{\dl 
      c_i\dl A\dl\cb} \lf(\df{\dl^2\G_k}{\dl c\dl\cb}+R_{c,k}^T\ri)^{-1} 
    \;\;. 
\end{split}\end{equation}
The other two terms are treated alike. The ghost flow equation then 
reads (using $R_{c,k}^T=R_{c,k}$) 
\begin{equation}\label{full ghost flow formula}\begin{split} 
    \fc{\dl^2\dot{\G}_k}{\dl\cb_j\dl c_i} =&\Tr \;\dot{R}_{A,k} 
    \lf(\df{\dl^2\G_k}{\dl A\dl A}+R_{A,k}\ri)^{-1} 
    \df{\dl^3\G_k}{\dl\cb_j \dl A\dl c} \lf(-\df{\dl^2\G_k}{\dl\cb\dl 
      c}+R_{c,k}\ri)^{-1} 
    \df{\dl^3\G_k}{\dl c_i\dl\cb\dl A} \lf(\df{\dl^2\G_k}{\dl A\dl A}+R_{A,k}\ri)^{-1} \\ 
    &+\Tr \;\dot{R}_{c,k} \lf(-\df{\dl^2\G_k}{\dl\cb\dl c} 
    +R_{c,k}\ri)^{-1} \df{\dl^3\G_k}{\dl c_i\dl\cb\dl A} 
    \lf(\df{\dl^2\G_k}{\dl A\dl A} +R_{A,k}\ri)^{-1} 
    \df{\dl^3\G_k}{\dl\cb_j\dl A\dl c} \lf(-\df{\dl^2\G_k}{\dl\cb\dl 
      c} 
    +R_{c,k}\ri)^{-1} \\ 
    &-\fc{1}{2}\,\Tr \;\dot{R}_{A,k} \lf(\df{\dl^2\G_k}{\dl A\dl 
      A}+R_{A,k}\ri)^{-1} \df{\dl^4\G_k}{\dl\cb_j\dl c_i\dl A\dl A} 
    \lf(\df{\dl^2\G_k}{\dl A\dl A} 
    +R_{A,k}\ri)^{-1} \\ 
    &-\Tr \;\dot{R}_{c,k} \lf(-\df{\dl^2\G_k}{\dl\cb\dl 
      c}+R_{c,k}\ri)^{-1} \df{\dl^4\G_k}{\dl\cb_j\dl c_i\dl \cb\dl c} 
    \lf(-\df{\dl^2\G_k}{\dl\cb\dl c}+R_{c,k}\ri)^{-1} \qquad . 
\end{split}\end{equation} 
In much the same way, the gluon flow equation is deduced from 
Eq. (\ref{flow of action}), and the result is 
\begin{equation}\label{full gluon flow formula}\begin{split} 
    \fc{\dl^2\dot{\G}_k}{\dl A_j\dl A_i} =& \Tr \;\dot{R}_{A,k} 
    \lf(\df{\dl^2\G_k}{\dl A\dl A}+R_{A,k}\ri)^{-1} \df{\dl^3\G_k}{\dl 
      A_j\dl A\dl A} \lf(\df{\dl^2\G_k}{\dl A\dl A}+ R_{A,k}\ri)^{-1} 
    \df{\dl^3\G_k}{\dl A_i\dl A\dl A} 
    \lf(\df{\dl^2\G_k}{\dl A\dl A}+R_{A,k}\ri)^{-1} \\ 
    &-\Tr \;\dot{R}_{c,k} \lf(-\df{\dl^2\G_k}{\dl\cb\dl c}+ 
    R_{c,k}\ri)^{-1} \df{\dl^3\G_k}{\dl A_j\dl\cb\dl c} 
    \lf(-\df{\dl^2\G_k}{\dl \cb\dl c} +R_{c,k}\ri)^{-1} 
    \df{\dl^3\G_k}{\dl A_i\dl\cb\dl c} \lf(-\df{\dl^2\G_k}{\dl\cb\dl 
      c} 
    +R_{c,k}\ri)^{-1} \\ 
    &-\Tr \;\dot{R}_{c,k} \lf(-\df{\dl^2\G_k}{\dl\cb\dl 
      c}+R_{c,k}\ri)^{-1} \df{\dl^3\G_k}{\dl A_i\dl\cb\dl c} 
    \lf(-\df{\dl^2\G_k}{\dl \cb\dl c} +R_{c,k}\ri)^{-1} 
    \df{\dl^3\G_k}{\dl A_j\dl\cb\dl c} 
    \lf(-\df{\dl^2\G_k}{\dl\cb\dl c}+R_{c,k}\ri)^{-1} \\ 
    &-\fc{1}{2}\,\Tr \;\dot{R}_{A,k} \lf(\df{\dl^2\G_k}{\dl A\dl A} 
    +R_{A,k}\ri)^{-1} \df{\dl^4\G_k}{\dl A_j\dl A_i\dl A\dl A} 
    \lf(\df{\dl^2\G_k}{\dl A\dl A}+R_{A,k}\ri)^{-1} \\ 
    &-\Tr \;\dot{R}_{c,k} \lf(-\df{\dl^2\G_k}{\dl\cb\dl 
      c}+R_{c,k}\ri)^{-1} \df{\dl^4\G_k}{\dl A_j\dl A_i\dl \cb\dl c} 
    \lf(-\df{\dl^2\G_k}{\dl\cb\dl c}+R_{c,k}\ri)^{-1} \;\;. 
\end{split}\end{equation} 
These equations are represented diagrammatically in Figs.
\ref{gluon full flow}, \ref{ghost full flow}.

\section{\protect\label{contractions} Explicit form of the flow equations} 
 
In this appendix we will use the parametrization of sect. \ref{parameterisation} in order 
to bring the truncated flow equations represented in Figs. 
\ref{gluon trunc flow}, \ref{ghost trunc flow} into their final form.
 
As the ghost two-point function and the ghost regulator are both 
diagonal in color space and momentum space, they can easily be 
inverted, yielding (see Eqs. (\ref{regulator}), (\ref{decomposition 
  ghost}), (\ref{eq:barGkC}), (\ref{eq:GkC}) ) 
\begin{equation}\begin{split} 
    \lf[\lf(- \fc{\dl^2\G_k}{\dl \cb\dl c} + 
    R_{c,k}\ri)^{-1}\ri]^{ab}_{\fp\fq} &= \dl^{ab} \lf[g \, p^2/d_k(p) 
    + R_{c,k}(p)\ri]^{-1} (2\pi)^3\dl^3(\fp+\fq) \\ 
    &= \dl^{ab}  \fc{1}{g}\bar{G}_{c,k}(p)(2\pi)^3\dl^3(\fp+\fq) \;\;. 
\end{split}\end{equation} 
A similar formula holds for the gluon two-point function and 
the gluon regulator, which are invertible in the transverse subspace 
(see Eqs. (\ref{regulator}), (\ref{decomposition gluon}), (\ref{eq:GkA})): 
\begin{equation} 
  \lf[\lf(\fc{\dl^2\G_k}{\dl A\dl A} + R_{A,k}\ri)^{-1}\ri]^{ab}_{ij,\fp\fq}  
  = \dl^{ab}t_{ij}(\fp) G_{A,k}(p)(2\pi)^3\dl^3(\fp+\fq) \;\;. 
\end{equation} 
Therefore, the truncated gluon flow equation, Fig. \ref{gluon trunc flow}, see also Eq. (\ref{full gluon flow formula}), reduces to 
\begin{equation}\begin{split} 
    2\dl^{ab}t_{ij}(\fp) \dot{\om}_k(p) & (2\pi)^3\dl^3 (\fp+\fq) = \\ 
    -\int \frac{d^3[p_{1\ldots 6}]}{(2\pi)^{18}}\; 
    &\dl^{cd}\dot{\bar{R}}_{c,k}(p_1)(2\pi)^3\dl^3(\fp_1-\fp_2) 
    \;\dl^{de}\bar{G}_{c,k}(p_2)(2\pi)^3\dl^3(\fp_2-\fp_3) \\ 
    &\times{} f^{eaf}t_{im}(\fp)(ip_{4,m})(2\pi)^3\dl^3(\fp+\fp_3-\fp_4) 
    \;\dl^{fg}\bar{G}_{c,k}(p_4)(2\pi)^3\dl^3(\fp_4-\fp_5) \\ 
    &\times{} f^{gbh}t_{jl}(\fq)(ip_{6,l})(2\pi)^3\dl^3(\fq+\fp_5-\fp_6) 
    \;\dl^{hc}\bar{G}_{c,k}(p_6)(2\pi)^3\dl^3(\fp_6-\fp_1) + 
    (i \leftrightarrow j) \;\;,
\end{split}\end{equation}
and by carrying out index contractions and integrations we obtain the flow equation for $\om_k$: 
\begin{equation} 
  \partial_t{\om}_k(p) = - \fc{N_c}{2}\int \frac{d^3 q}{(2\pi)^3} \; 
  \lf(\bar{G}_{c,k}\dot{\bar{R}}_{c,k} \bar{G}_{c,k}\ri)(q) \bar{G}_{c,k}(|\fp+\fq|) 
  \;q^2(1-(\hat{\fp}\cdot\hat{\fq})^2) \;\;. 
\end{equation} 
In much the same way we treat the truncated ghost flow equation, 
Fig. \ref{ghost trunc flow}, see also Eq. (\ref{full ghost flow formula}), to get 
\begin{equation}\begin{split} 
    -\dl^{ab} \partial_t d_k^{-1}(p)  (2\pi)^3 & \dl^3  (\fp+\fq) = \\ 
    \int \frac{d^3[p_{1\ldots 6}]}{(2\pi)^{18}} \; 
    &\dl^{cd}t_{ij}(\fp_1)\dot{R}_{A,k}(p_1)(2\pi)^3\dl^3(\fp_1-\fp_2) 
    \;\dl^{de}t_{jh}(\fp_2)G_{A,k}(p_2)(2\pi)^3\dl^3(\fp_2-\fp_3) \\ 
    &\times{} f^{fea}t_{lh}(\fp_3)(-ip_{4,l})(2\pi)^3\dl^3(\fp+\fp_3-\fp_4) 
    \;\dl^{fg}\bar{G}_{c,k}(p_4)(2\pi)^3\dl^3(\fp_4-\fp_5) \\ 
    &\times{} f^{bhg}t_{mn}(\fp_6)(iq_m)(2\pi)^3\dl^3(\fq+\fp_5-\fp_6) 
    \;\dl^{hc}t_{ni}(\fp_6)G_{A,k}(p_6)(2\pi)^3\dl^3(\fp_6-\fp_1) \\ 
    + \int \frac{d^3[p_{1\ldots 6}]}{(2\pi)^{18}} \; 
    &\dl^{hc}\dot{\bar{R}}_{c,k}(p_1)(2\pi)^3\dl^3(\fp_1-\fp_2) 
    \;\dl^{cd}\bar{G}_{c,k}(p_2)(2\pi)^3\dl^3(\fp_2-\fp_3) \\ 
    &\times{} f^{bed}t_{mi}(\fp_4)(iq_m)(2\pi)^3\dl^3(\fq+\fp_3-\fp_4) 
    \;\dl^{ef}t_{ij}(\fp_4)G_{A,k}(p_4)(2\pi)^3\dl^3(\fp_4-\fp_5) \\ 
    &\times{} f^{gfa}t_{hj}(\fp_5)(-ip_{6,h})(2\pi)^3\dl^3(\fp+\fp_5-\fp_6) 
    \;\dl^{gh}\bar{G}_{c,k}(p_6)(2\pi)^3\dl^3(\fp_6-\fp_1) 
    \;\;, 
\end{split}\end{equation} 
which, after receiving the same treatment as the gluon equation above, 
finally turns into, 
\begin{equation}\begin{split} 
    \partial_t d_k^{-1}(p) = N_c \, p^2 \int \frac{d^3 q}{(2\pi)^3} \; 
    &\bigg[\lf(G_{A,k}\dot{R}_{A,k}G_{A,k}\ri)(q) \bar{G}_{c,k}(|\fp+\fq|) (1-(\hat{\fp}\cdot\hat{\fq})^2) \\ 
    &+\lf(\bar{G}_{c,k}\dot{\bar{R}}_{c,k} \bar{G}_{c,k}\ri)(q) G_{A,k} 
    (|\fp+\fq|) \;q^2 \; 
    \fc{1-(\hat{\fp}\cdot\hat{\fq})^2}{(\fp+\fq)^2} \bigg] \;\;. 
\end{split}\end{equation} 
\end{widetext}

\section{Numerical Computation}\label{numerics} 
 
Below we give some details of the numerical method used for the 
solution of the flow equations.  In order to solve the flow equations 
numerically, the functions $\om(p)$ and 
$d(p)$ of section \ref{analytical integration} as well as $\om_k(p)$ 
and $d_k(p)$ of section \ref{numerical integration} are represented by 
means of Chebyshev polynomials on suitably sized momentum ranges, 
spanning up to nine orders of magnitude. We use a logarithmic momentum 
scale to sample the behavior of the functions equally well on all 
orders of magnitude; the function values are also represented 
logarithmically. In the case of the flow functions $\om_k(p)$ and 
$d_k(p)$, which possess two momentum arguments, we first calculate the 
coefficients along the $p$-direction at constant values of $k$ which are the Chebyshev nodes of the $k$-range, then 
determine the coefficients of the resulting functions in 
$k$-direction for each $p$-coefficient. About 130 Chebyshev nodes in each direction have been 
used. The momentum integrals, for the loop as well as for the flow (in 
section \ref{numerical integration}), have been calculated using the 
Gauss-Legendre method with about 70 nodes. As for the Chebyshev 
nodes, the Gauss-Legendre nodes for the momentum integrals have been 
calculated for a logarithmic momentum scale in order to appropriately 
sample the IR behavior. The loop integrals have been confined to a 
range of one order of magnitude around the cutoff momentum $k$ outside 
of which there is virtually no contribution to the integral, owing to 
the regulator functions employed, see Eq. (\ref{explicit shape}). 
 
 In the loop integrals, the flow functions need to be evaluated also 
 outside the Chebyshev representation range of the external 
 momentum $p$. But the IR extrapolation of the flow functions requires no 
 extra assumptions about their IR behavior: as long as $k_{min}$ never 
 reaches the lower boundary of the $p$-range, the functions can be 
 simply chosen to be constant beyond this boundary, as clearly seen in Fig. \ref{ghost full flow 3dim pic}. For their continuation in the UV, a 
 power law has been fitted to their behavior in the UV region of the 
 representation range, which is $\om_k(p) \sim p$ and $d_k(p) \sim 
 const.$ As for the $k$-range, 
 the functions are never evaluated outside the Chebyshev range anyway, 
 so no extrapolation is necessary.
 
The sets of equations (\ref{int flow om symbolical}), (\ref{int flow d symbolical}) as well as 
(\ref{ghost simple}), (\ref{gluon simple}) have been evaluated 
iteratively: starting from constant functions $\om$ and $d$, the 
r.h.s. of the equations are calculated; from the result, the initial 
values of the flow, $d_\Lambda$ and $\om_\Lambda$, are determined 
following the procedures described in the next paragraph, giving temporary results 
$\om^{tmp}\,,\, d^{tmp}$. To improve the convergence behavior, we use 
a relaxation method to determine the final result of the $n$-th 
iteration as $\om_k^{(n)}(p) = r\om_k^{tmp}(p) + (1-r) 
\om_k^{(n-1)}(p)$ for each Chebyshev node $(k,p)$, likewise for 
$d_k(p)$. Values of $r \in [0.1,0.5]$ have been used, depending on how 
much the functions change from one iteration to the next. This yields 
the new functions $\om^{(n)}$ and $d^{(n)}$ which are then fed back 
into the r.h.s. of the equations as input to the $(n+1)$-th step of the iteration. This iteration is repeated until convergence is reached.

To determine the constant $d_\Lambda(p) \equiv d_\Lambda$ in Eq. (\ref{int flow d symbolical}), we demand that $d_{k_{min}}(p)$ fulfill a power law for $p$ in the IR, $d^{-1}_{k_{min}}(p) \sim p^\beta$. To do so, we observe that for a power law $f(p) = p^\beta$, the expression
\begin{equation}
p \frac{d}{dp}\ln f(p) = \beta
\end{equation}
yields a constant value, i.e., the exponent. Let $g(p)$ be the r.h.s. of Eq. (\ref{int flow d symbolical}) (with $k=k_{min}$). We demand that
\begin{equation}
\frac{d}{dp}\lf(p \frac{d}{dp}\ln \lf(g(p) + d_\Lambda^{-1}\ri)\ri) \stackrel{!}{=} 0
\end{equation}
and solve for $d_\Lambda^{-1}$. A least squares fit of a constant function to the expression obtained (which will in general not yet be constant) in the IR  gives the optimal value for $d_\Lambda^{-1}$ in order to achieve a power law behavior for $d^{-1}_{k_{min}}(p)$ in the IR region. This is done during each iteration step as described in the previous paragraph. In this way, we can impose a power law behavior on the ghost form factor without dictating its exponent.
Concerning the Eqs. (\ref{int flow om symbolical}) (with $k=k_{min}$) and (\ref{gluon simple}), we fit the expression $-a + bp$ to the r.h.s. in the UV and use $\om_\Lambda(p) = a+p$ to achieve $\om_{k=k_{min}}(p)|_{p\rightarrow\Lambda} \sim p$ for Eq. (\ref{int flow om symbolical}) and $\om_0(p)|_{p\rightarrow\Lambda} \sim p$ for Eq. (\ref{gluon simple}) which is an expression of asymptotic freedom. For Eq. (\ref{ghost simple}) the determination of the initial conditions is not a numerical issue anyway, see Eq. (\ref{ghost simple subtracted}).

This method allows for a systematic, simultaneous determination of the 
solution together with originally unknown initial conditions to 
accomplish that the solution at $k=k_{min}$ fulfill certain 
properties. Solving the differential equations in the 
differential form would require a trial-and-error search for the initial conditions 
to find a solution with the specified properties.


\begin{thebibliography}{99} 
 


\bibitem{Fischer:2008uz} 
  C.~S.~Fischer, A.~Maas and J.~M.~Pawlowski, 
  Annals Phys.\  {\bf 324} (2009) 2408 
  [arXiv:0810.1987 [hep-ph]]. 

\bibitem{Alkofer:2000wg}
  R.\ Alkofer and L.\ von Smekal,
  Phys.\ Rept.\ {\bf 353} (2001) 281
  [arXiv:hep-ph/0007355].

\bibitem{Fischer:2006ub}
  C.\ S.\ Fischer,
  J.\ Phys.\ G {\bf 32}, R253 (2006)
  [arXiv:hep-ph/0605173].


\bibitem{vonSmekal:2008ws} 
  L.~von Smekal, 
  arXiv:0812.0654 [hep-th]. 


\bibitem{Binosi:2009qm}
  D.~Binosi and J.~Papavassiliou,
  Phys.\ Rept.\  {\bf 479} (2009) 1
  [arXiv:0909.2536 [hep-ph]].

\bibitem{Boucaud:2008ky}
  P.~Boucaud, J.~P.~Leroy, A.~Le Yaouanc, J.~Micheli, O.~Pene and J.~Rodriguez-Quintero,
  JHEP {\bf 0806} (2008) 099
  [arXiv:0803.2161 [hep-ph]].
 
\bibitem{Litim:1998nf}
  D.~F.~Litim and J.~M.~Pawlowski,
  arXiv:hep-th/9901063.



\bibitem{Pawlowski:2005xe} 
  J.~M.~Pawlowski, 
  Annals Phys.\  {\bf 322} (2007) 2831 
  [arXiv:hep-th/0512261]. 

\bibitem{Gies:2006wv}
  H.~Gies,
  arXiv:hep-ph/0611146.
 
\bibitem{Ellwanger:1995qf} 
  U.~Ellwanger, M.~Hirsch and A.~Weber, 
  Z.\ Phys.\  C {\bf 69} (1996) 687 
  [arXiv:hep-th/9506019]. 


\bibitem{Zwanziger:1998ez} 
  D.~Zwanziger, 
  Nucl.\ Phys.\  B {\bf 518} (1998) 237.\\ 
L.~Baulieu and D.~Zwanziger, 
  Nucl.\ Phys.\  B {\bf 548}, 527 (1999) 
  [arXiv:hep-th/9807024]. 
 

\bibitem{Watson:2006yq} 
  P.~Watson and H.~Reinhardt, 
  Phys.\ Rev.\  D {\bf 75}, 045021 (2007) 
  [arXiv:hep-th/0612114].\\ 
P.~Watson and H.~Reinhardt, 
  Phys.\ Rev.\  D {\bf 76}, 125016 (2007) 
  [arXiv:0709.0140 [hep-th]].\\ 
P.~Watson and H.~Reinhardt, 
  Phys.\ Rev.\  D {\bf 77}, 025030 (2008) 
  [arXiv:0709.3963 [hep-th]]. 
 


\bibitem{Alkofer:2009dm}
  R.~Alkofer, A.~Maas and D.~Zwanziger,
  Few Body Syst.\  {\bf 47} (2010) 73
  [arXiv:0905.4594 [hep-ph]].

\bibitem{Szczepaniak:2001rg} 
  A.~P.~Szczepaniak and E.~S.~Swanson, 
  Phys.\ Rev.\  D {\bf 65}, 025012 (2002) 
  [arXiv:hep-ph/0107078].\\ 
      A.~P.~Szczepaniak, 
        Phys.\ Rev.\  D {\bf 69}, 074031 (2004) 
        [arXiv:hep-ph/0306030]. 
 
 
\bibitem{Feuchter:2004mk} 
  C.~Feuchter and H.~Reinhardt, 
  Phys.\ Rev.\  D {\bf 70} (2004) 105021 
  [arXiv:hep-th/0408236], 
[arXiv:hep-th/0402106] 
 
\bibitem{Reinhardt:2004mm} 
  H.~Reinhardt and C.~Feuchter, 
  Phys.\ Rev.\  D {\bf 71}, 105002 (2005) 
  [arXiv:hep-th/0408237]. 
 
\bibitem{Schleifenbaum:2006bq} 
  W.~Schleifenbaum, M.~Leder and H.~Reinhardt, 
  Phys.\ Rev.\  D {\bf 73}, 125019 (2006) 
  [arXiv:hep-th/0605115]. 
 
      \bibitem{Epple:2006hv} 
        D.~Epple, H.~Reinhardt and W.~Schleifenbaum, 
        Phys.\ Rev.\  D {\bf 75}, 045011 (2007) 
        [arXiv:hep-th/0612241]. 
 
\bibitem{Epple:2007ut} 
  D.~Epple, H.~Reinhardt, W.~Schleifenbaum and A.~P.~Szczepaniak, 
  Phys.\ Rev.\  D {\bf 77}, 085007 (2008) 
  [arXiv:0712.3694 [hep-th]]. 
 
\bibitem{Reinhardt:2008ax}
  H.~Reinhardt, D.~Campagnari, D.~Epple, M.~Leder, M.~Pak and W.~Schleifenbaum,
  arXiv:0807.4635 [hep-th].

\bibitem{Reinhardt:2008ek}
  H.~Reinhardt,
  Phys.\ Rev.\ Lett.\  {\bf 101} (2008) 061602
  [arXiv:0803.0504 [hep-th]].

 
\bibitem{Wetterich:1992yh} 
  C.~Wetterich, 
  Phys.\ Lett.\  B {\bf 301} (1993) 90. 
 
\bibitem{Burgio:2009xp}
  G.~Burgio, M.~Quandt and H.~Reinhardt,
  Phys.\ Rev.\  D {\bf 81} (2010) 074502
  [arXiv:0911.5101 [hep-lat]].
 
 
 
\bibitem{Fischer:2009tn} 
  C.~S.~Fischer and J.~M.~Pawlowski, 
  Phys.\ Rev.\  D {\bf 80} (2009) 025023 
  [arXiv:0903.2193 [hep-th]]; 

  Phys.\ Rev.\  D {\bf 75} (2007) 025012 
  [arXiv:hep-th/0609009]. 



\bibitem{Alkofer:2008jy}
  R.~Alkofer, M.~Q.~Huber and K.~Schwenzer,
  Phys.\ Rev.\  D {\bf 81} (2010) 105010
  [arXiv:0801.2762 [hep-th]]; 
  arXiv:0904.1873 [hep-th].

\bibitem{Lerche:2002ep}
  C.~Lerche and L.~von Smekal,
  Phys.\ Rev.\  D {\bf 65} (2002) 125006
  [arXiv:hep-ph/0202194].


\bibitem{RX1} 
D.~Campagnari, A.~Weber, H.~Reinhardt, F.~Astorga and W.~Schleifenbaum, 
arXiv:0910.4548 [hep-th]. 



\bibitem{Zwa92} 
D.~Zwanziger,  
Nucl. Phys. {\bf B364}, 127 (1991); 
Nucl. Phys. {\bf B485}, 185 (1997); 
Phys. Rev. D {\bf 70}, 094034 (2004). 


\bibitem{Tay71} 
J.~C.~Taylor, 
Nucl.\ Phys.\ {\bf B33}, 436 (1971); 
W.~Marciano and H.~Pagels, 
Phys.\ Rept.\ {\bf 36}, 137 (1978). 


\bibitem{CMM04} 
A.~Cucchieri, T.~Mendes and A.~Mihara,  
JHEP {\bf 12}, 012 (2004); 
A.~Sternbeck, E.~M.~Ilgenfritz, M.~Muller-Preussker and A.~Schiller, 
Nucl.\ Phys.\ B Proc.\ Suppl.\ {\bf153}, 185 (2006).


\bibitem{CRW09} 
D.~Campagnari, H.~Reinhardt and A.~Weber,  
Phys.\ Rev.\ D {\bf 80}, 025005 (2009)  
[arXiv:0904.3490 [hep-th]]. 

\bibitem{Fischer:2005qe}
  C.~S.~Fischer and D.~Zwanziger,
  Phys.\ Rev.\  D {\bf 72} (2005) 054005
  [arXiv:hep-ph/0504244].

\bibitem{Ellwanger:1994iz} 
  U.~Ellwanger, 
  %
  Phys.\ Lett.\ B {\bf 335} (1994) 364. 

\bibitem{Freire:2000bq} 
  F.~Freire, D.~F.~Litim and J.~M.~Pawlowski, 
  Phys.\ Lett.\  B {\bf 495} (2000) 256 
  [arXiv:hep-th/0009110]. 


\bibitem{Pawlowski} 
  J.~M.~Pawlowski,  in preparation. 


\bibitem{Pawlowski:2003hq} 
  J.~M.~Pawlowski, D.~F.~Litim, S.~Nedelko and L.~von Smekal, 
  Phys.\ Rev.\ Lett.\  {\bf 93} (2004) 152002 
  [arXiv:hep-th/0312324]. 
 



\bibitem{Litim:2000ci} 
  D.~F.~Litim, 
  %
  Phys.\ Lett.\ B {\bf 486} (2000) 92;
  Int.\ J.\ Mod.\ Phys.\ A {\bf 16} (2001) 2081. 
 


\bibitem{Braun:2007bx} 
  J.~Braun, H.~Gies and J.~M.~Pawlowski, 
  Phys.\ Lett.\  B {\bf 684} (2010) 262 
  [arXiv:0708.2413 [hep-th]]. 
 
 
\bibitem{Braun:2009gm}
  J.~Braun, L.~M.~Haas, F.~Marhauser and J.~M.~Pawlowski,
  arXiv:0908.0008 [hep-ph].

\end{thebibliography}
\end{document}